\documentclass[12pt]{amsart}
\usepackage{amsmath,amssymb,hyperref,enumitem,graphicx}
\usepackage[sort&compress,numbers]{natbib}

\hypersetup{colorlinks=true,allcolors=blue}
\usepackage[all]{hypcap}

\usepackage[foot]{amsaddr}
\usepackage[sc]{mathpazo}

\newcommand\ack{\subsection*{Acknowledgment}}

\DeclareMathAlphabet\mathsfbi{T1}{phv}{b}{it}

\numberwithin{equation}{section}

\newcommand\BV{\boldsymbol} 
\newcommand\BM{\mathsfbi} 

\newcommand\dif{\:\!\mathrm{d}}
\newcommand\deriv[2]{\frac{\mathrm{d} #1}{\mathrm{d} #2}}
\newcommand\parderiv[2]{\frac{\partial #1}{\partial #2}}

\newcommand\trace{\mathrm{tr}}

\newcommand\RR{\mathbb R}

\newcommand\myatop[2]{\genfrac{}{}{0pt}{}{#1}{#2}}

\def \Mach{\textit{Ma}}
\def \Rey {\textit{Re}}
\def \Nu {\mathcal V}

\begin{document}

\author[Rafail V. Abramov]{Rafail V. Abramov}

\address{Department of Mathematics, Statistics and Computer Science,
University of Illinois at Chicago, 851 S. Morgan st., Chicago, IL 60607}

\email{abramov@uic.edu}

\title{Formation of turbulence via an interaction potential}

\begin{abstract}
In a recent work, we proposed a hypothesis that the turbulence in
gases could be produced by particles interacting via a potential --
for example, the interatomic potential at short ranges, and the
electrostatic potential at long ranges. Here, we examine the proposed
mechanics of turbulence formation in a simple model of two particles,
which interact solely via a potential. Following the kinetic theory
approach, we derive a hierarchy of the velocity moment transport
equations, and then truncate it via a novel closure based on the high
Reynolds number condition. While standard closures of the velocity
moment hierarchy of the Boltzmann equation lead to the compressible
Euler and Navier--Stokes systems of equations, our closure leads to a
transport equation for the velocity alone, which is driven by the
potential forcing. Starting from a large scale laminar shear flow, we
numerically simulate the solutions of our velocity transport equation
for the electrostatic, gravity, Thomas--Fermi and Lennard-Jones
potentials, as well as the Vlasov-type large scale mean field
potential. In all studied scenarios, the time-averaged Fourier spectra
of the kinetic energy clearly exhibit Kolmogorov's ``five-thirds''
power decay rate.
\end{abstract}

\maketitle

\section{Introduction}

In our recent work \citep{Abr20}, we proposed a hypothesis that the
turbulence in gases could be produced by their particles interacting
via a potential. The reason why we proposed such an unusual hypothesis
is the following. We investigated the dynamical system consisting of
many particles interacting via a potential, without reducing it to a
single-particle equation as typically done in the conventional kinetic
theory. We found that, due to the presence of the potential, a strong
large scale flow creates the forcing in the three-dimensional bundles
of the full coordinate space, with each bundle belonging to a pair of
particles. At the same time, in the Boltzmann
equation~\citep{Bol,CerIllPul}, the potential interactions are
replaced with the collision integral, which results in the absence of
the potential forcing terms in the standard equations of fluid
mechanics, such as the Euler~\citep{Bat} or
Navier--Stokes~\citep{Gols} equations.

However, we noted that the direct observations and measurements of a
turbulent fluid can register some bulk properties of the induced flow
in these particle-pair bundles, such as the power scaling of the
Fourier transform of the kinetic energy of the flow. In our
work~\citep{Abr20}, we made crude estimates of the power scaling of
the kinetic energy in the inertial range, induced by a strong large
scale flow via the interaction potential. Remarkably, we found our
crude estimates to be consistent with direct measurements and
observations \citep{BucVel,NasGag}.

Our results \citep{Abr20} further motivated us to look for a more
detailed explanation of how the turbulence could be induced by a
strong large scale flow via an interaction potential. In the current
work, we develop and implement a simple fluid mechanical model of
behavior of a pair of particles, which interact solely via a
potential. While we recognize that, in the real world natural
phenomena, the dynamics are much more complex with multiple types of
interactions, the primary goal of the current work is to investigate
whether an interaction potential alone by itself can produce flow
structures with power decay of the Fourier transforms of dynamical
variables. If needed, an extension onto many particles can be made in
a standard fashion via the
Bogoliubov--Born--Green--Kirkwood--Yvon~\citep{Bog,BorGre,Kir}
hierarchy approach.

The paper is organized as follows. In
Section~\ref{sec:particle_dynamics}, we start with the equations of
motion for a system of two particles, which interact via a
potential. For this system, we compute the Liouville equation, the
thermodynamic equilibrium state of the system, and show that a generic
solution preserves the ``distance'' to the equilibrium state in the
sense of any of the R\'enyi metrics \citep{Ren}, which is similar to
the effect of Boltzmann's $H$-theorem~\citep{Bol} for hard-sphere
gases. We change the variables from the coordinate and velocity of
each individual particle to those of the center of mass of the system
and the difference between the particles. We then exclude the motion
of the center of mass from the dynamics, which leads to the Liouville
equation for the turbulent variables (that is, those which represent
the differences in the coordinates and velocities between the
particles).  Alternatively, we integrate out the second particle, and
arrive at the Vlasov equation \citep{Vla} for a single particle, which
has the same form as the Liouville equation for the turbulent
variables.

In Section \ref{sec:turbulent_velocity}, we formulate a hierarchy of
the transport equations for the velocity moments of the turbulent
Liouville or Vlasov equations. However, due to the fact that the
potential forcing replaces the usual Boltzmann collision integral, a
different closure must be used to truncate the moment hierarchy. Here,
we introduce a novel closure based on the high Reynolds number
condition, which leads to a single equation for the turbulent
velocity, driven by the potential forcing. This corresponds well with
the fact that the observed turbulence appears to have largely
universal behavior across a variety of different media. However, due
to the simplicity of our model, the turbulent velocity equation lacks
dissipation, and thus its solutions are meaningful on a limited time
scale.

In Section \ref{sec:numerical} we show the results of numerical
simulations of the turbulent velocity equation, which all start with a
large scale laminar shear flow as the initial condition, and are
forced by different types of the interaction potential. We examine the
scenarios for the following interaction potentials: electrostatic,
gravitational, Thomas--Fermi~\citep{Tho,Fer},
Lennard-Jones~\citep{Len}, as well as the Vlasov-type large scale mean
field potential. In each scenario, we discover the regime of secular
growth which precludes the exponential blow-up (the latter due to the
lack of dissipation). For all scenarios, in this secular growth
regime, the time-averaged Fourier transforms of the kinetic energy of
the flow decay as the negative five-thirds power of the wavenumber,
which matches the Kolmogorov turbulence
spectrum~\citep{Kol41a,Kol41b,Kol41c,Obu41,Obu49,Obu62}. The results
of this work are summarized in Section~\ref{sec:summary}. The
generalizations onto multiparticle systems are sketched in
Appendices~\ref{app:multiparticle}, \ref{app:vlasov} and
\ref{app:closure_pair}.

\section{Particle dynamics}
\label{sec:particle_dynamics}

In our attempt to uncover the origins of the turbulent kinetic energy
spectra, we follow the general approach of kinetic theory, starting
with the elementary evolution mechanics of particles which interact
solely via a potential. First, we consider a simple model of only two
interacting particles, thereby avoiding the multiparticle closure
problem, and derive the Liouville equation for this pair of particles.
Second, we derive the Vlasov equation for one of the particles in the
pair, by using a simple closure for another particle. The
generalizations of what is presented here onto a multiparticle set-up
are given in Appendices \ref{app:multiparticle}, \ref{app:vlasov} and
\ref{app:closure_pair}.

\subsection{Two particle dynamics}

Let the two identical particles, with coordinates $\BV x_1$
and $\BV x_2$, and velocities $\BV v_1$ and $\BV v_2$, respectively,
interact via a potential $\phi(r)$. The equations of motion for these
two particles are given via
\begin{subequations}
\label{eq:dyn_sys}
\begin{equation}
\deriv{\BV x_1}t=\BV v_1,\qquad\deriv{\BV v_1}t=-\parderiv{}{\BV x_1}
\phi(\|\BV x_2-\BV x_1\|),
\end{equation}
\begin{equation}
\deriv{\BV x_2}t=\BV v_2,\qquad\deriv{\BV v_2}t=-\parderiv{}{\BV x_2}
\phi(\|\BV x_2-\BV x_1\|).
\end{equation}
\end{subequations}
Here we make two assumptions, which simplify further manipulations.
First, we assume that the coordinate domain is of finite volume, but
has no discernible boundaries (e.g. a periodic cube). Second, we
assume that the potential $\phi(r)$ does not have a singularity at
zero, although it may still peak strongly as $r\to 0$ to model either
repulsion or attraction, whichever is needed in the context of the
problem.

It is easy to see that the dynamics in \eqref{eq:dyn_sys} preserve the
momentum and energy of the system of two particles:
\begin{equation}
\BV v_1+\BV v_2=\text{const},\qquad\|\BV v_1\|^2+\|\BV v_2\|^2+
2\phi(\|\BV x_2-\BV x_1\|)=\text{const}.
\end{equation}
For a given value of the momentum, it is always possible to choose the
inertial reference frame in which the momentum becomes zero (the
so-called Galilean shift). Thus, without much loss of generality, we
will further assume that the total momentum of the system of the two
particles is zero.

\subsection{The Liouville equation}

Let $f(t,\BV x_1,\BV v_1,\BV x_2,\BV v_2)$ denote the distribution
density of states of \eqref{eq:dyn_sys}. Then, the transport equation
for $f$, known as the Liouville equation, is given via
\begin{equation}
\label{eq:liouville_2}
\parderiv ft+\BV v_1\cdot\parderiv f{\BV x_1}+\BV v_2\cdot\parderiv
f{\BV x_2}=\parderiv{}{\BV x_1}\phi(\|\BV x_2-\BV x_1\|)\cdot\parderiv
f{\BV v_1}+\parderiv{}{\BV x_2}\phi(\|\BV x_2-\BV x_1\|)\cdot\parderiv
f{\BV v_2}.
\end{equation}
One can verify that any suitable function of the form
\begin{equation}
\label{eq:f_0}
f_0(\BV x_1,\BV v_1,\BV x_2,\BV v_2)=g\big(\|\BV v_1\|^2+\|\BV
v_2\|^2+2\phi(\|\BV x_2-\BV x_1\|)\big)
\end{equation}
is a steady state. Among all such states, the canonical Gibbs state is
given via
\begin{equation}
\label{eq:f_G}
f_G(\BV x_1,\BV v_1,\BV x_2,\BV v_2)=\frac 1{(2\pi\theta_0)^3Z}
\exp\left(-\frac{\|\BV v_1\|^2+\|\BV v_2\|^2+2\phi(\|\BV x_2-\BV
  x_1\|)}{2\theta_0}\right),
\end{equation}
where $\theta_0$ is the kinetic temperature of the system, and $Z$ is
the spatial normalization constant:
\begin{equation}
Z=\int e^{-\phi(\|\BV x_2-\BV x_1\|)/\theta_0}\dif\BV x_1\dif\BV x_2.
\end{equation}

\subsection{Preservation of the R\'enyi divergence}

It is important to note that the Liouville equation preserves the
family of R\'enyi divergences \citep{Ren} between a solution $f$ and a
steady state $f_0$. Indeed, let $\psi_1:\RR\to\RR$, $\psi_2:\RR\to\RR$
be two differentiable functions. In the absence of boundary effects,
any quantity of the form
\begin{equation}
\int\psi_1(f)\psi_2(f_0)\dif\BV x_1\dif\BV v_1\dif\BV x_2\dif\BV v_2
\end{equation}
is preserved in time. Indeed, the time derivative yields the following
chain of identities (the measure notations are omitted to save space):
\begin{multline}
\parderiv{}t\int\psi_1(f)\psi_2(f_0)=\int\psi_2(f_0)\parderiv{\psi_1
  (f)}t=\int\psi_2(f_0)\bigg(\parderiv\phi{(\BV x_1,\BV x_2)}\cdot
\parderiv{ \psi_1(f)}{(\BV v_1,\BV v_2)}-\\-(\BV v_1,\BV v_2)\cdot
\parderiv{\psi_1(f)}{(\BV x_1,\BV x_2)}\bigg)=\int\psi_1(f) \bigg((\BV
v_1,\BV v_2)\cdot\parderiv{\psi_2(f_0)}{(\BV x_1,\BV x_2)}-\parderiv
\phi{(\BV x_1,\BV x_2)}\cdot\parderiv{\psi_2(f_0)}{(\BV v_1,\BV v_2)}
\bigg)=0.
\end{multline}
In particular, taking $\psi_1(x)=x^\alpha$, $\psi_2(x)=x^{1-\alpha}$,
for some $\alpha>0$, demonstrates that the family of general R\'enyi
divergences is preserved in time:
\begin{equation}
\label{eq:renyi_conservation}
D_\alpha(f,f_0)=\frac 1{\alpha-1}\ln\int f^\alpha f_0^{1-\alpha}
\dif\BV x_1\dif\BV v_1\dif\BV x_2\dif\BV v_2=\text{const}.
\end{equation}
The Kullback--Leibler divergence \citep{KulLei} is a special case of
the R\'enyi divergence with $\alpha=1$.

The above result shows that, in the absence of external or boundary
effects, and irreversible interactions, the dynamical system in
\eqref{eq:dyn_sys} retains is initial ``distance'' to any steady state
$f_0$ in the sense of the R\'enyi metric
\eqref{eq:renyi_conservation}. In particular, if it starts near the
Gibbs equilibrium \eqref{eq:f_G}, then it will remain so throughout
its time evolution.

\subsection{Mean and turbulent variables}

Let us make the following change of variables:
\begin{equation}
\label{eq:mean_turbulent_variables}
\BV x=\BV x_2-\BV x_1,\qquad\BV v=\BV v_2-\BV v_1,\qquad\BV y=
\frac{\BV x_1+\BV x_2}2,\qquad\BV w=\frac{\BV v_1+\BV v_2}2.
\end{equation}
Here, $\BV y$ and $\BV w$ describe the motion of the center of mass of
the system (and thus can be viewed as the ``mean'' variables), while
$\BV x$ and $\BV v$ describe the relative motions of one particle with
respect to another, and thus are regarded as the ``turbulent''
variables.  In the new variables, the partial derivatives become
\begin{subequations}
\label{eq:mean_turbulent_derivatives}
\begin{equation}
\parderiv{}{\BV x_1}=-\parderiv{}{\BV x}+\frac 12 \parderiv{}{\BV y},
\qquad\parderiv{}{\BV v_1}=-\parderiv{}{\BV v}+\frac 12\parderiv{}{\BV
  w},
\end{equation}
\begin{equation}
\parderiv{}{\BV x_2}= \parderiv{}{\BV x}+\frac 12\parderiv{}{\BV y},
\qquad\parderiv{}{\BV v_2}=\parderiv{}{\BV v}+\frac 12\parderiv{}{\BV
  w}.
\end{equation}
\end{subequations}
Substituting the expressions above into the Liouville equation for
$f$, we arrive at
\begin{equation}
\parderiv ft+\BV v\cdot\parderiv f{\BV x}+\BV w\cdot\parderiv f{\BV y}
=2\parderiv{\phi(\|\BV x\|)}{\BV x}\cdot\parderiv f{\BV v}.
\end{equation}
Observe that the total momentum variable $\BV w$ is merely a constant
parameter, because there are no derivatives with respect to it. Since
we consider the dynamics of \eqref{eq:dyn_sys} with zero total
momentum, we have to set $\BV w=\BV 0$ above. This leads to the
following Liouville equation for $f$:
\begin{equation}
\label{eq:liouville}
\parderiv ft+\BV v\cdot\parderiv f{\BV x}=2\parderiv{\phi(\|\BV
  x\|)}{\BV x}\cdot\parderiv f{\BV v}.
\end{equation}
Observe that dependence of $f$ on $\BV y$ and $\BV w$ no longer
matters, and further we assume that $f$ is only a function of $t$,
$\BV x$ and $\BV v$. The general steady state of \eqref{eq:liouville}
is given via
\begin{equation}
\label{eq:f_0t}
f_0(\BV x,\BV v)=g\big(\|\BV v\|^2+4\phi(\|\BV x\|)\big),
\end{equation}
with the corresponding Gibbs state being
\begin{equation}
\label{eq:f_Gt}
f_G(\BV x,\BV v)=\frac 1{(4\pi\theta_0)^3Z}\exp\left(-\frac{\|\BV
  v\|^2+4\phi(\|\BV x\|)}{4\theta_0}\right),\qquad Z=\int
e^{-\phi(\|\BV x\|)/\theta_0}\dif\BV x.
\end{equation}
The reason why $\theta_0$ is multiplied by a factor of 4 in the
denominator (instead of the usual 2) is because
\begin{equation}
\|\BV v\|^2=\|\BV v_2-\BV v_1\|^2=\|\BV v_2-\BV v_1\|^2+\|\BV v_1+\BV
v_2\|^2=2(\|\BV v_1\|^2+\|\BV v_2\|^2),
\end{equation}
where we use the fact that the total momentum of the system is zero.
So, if $\theta_0$ is the energy per degree of freedom per particle,
then the corresponding ``temperature of the difference'' between the
two particles must be twice that value.

Clearly, the Liouville equation \eqref{eq:liouville} also preserves
the family of R\'enyi metrics in \eqref{eq:renyi_conservation}. This
justifies the standard fluid-mechanical treatment of the dynamics via
the velocity-moment hierarchy with a subsequent closure, which we
apply further below.

\subsection{Single particle dynamics (Vlasov equation)}

The two-particle Liouville equation in~\eqref{eq:liouville_2} can be
reduced to the single-particle Vlasov equation \citep{Vla} with the
help of a closure. Let us denote the first particle marginal density
via
\begin{equation}
f_1(\BV x_1,\BV v_1)=\int f(\BV x_1,\BV v_1,\BV x_2,\BV
v_2)\dif\BV x_2\dif\BV v_2.
\end{equation}
The transport equation for $f_1$ can be obtained by integrating the
two-particle Liouville equation in \eqref{eq:liouville_2} in $\dif\BV
x_2\dif\BV v_2$:
\begin{equation}
\label{eq:f_1}
\parderiv{f_1}t+\BV v_1\cdot\parderiv{f_1}{\BV x_1}=\int\parderiv{
}{\BV x_1}\phi(\|\BV x_2-\BV x_1\|)\cdot\parderiv f{\BV v_1}\dif\BV
x_2\dif\BV v_2.
\end{equation}
Observe that the right-hand side above still contains $f$, and we need
a suitable closure to approximate it in terms of $f_1$. Here, we can
use the same principle as we did in our earlier work \citep{Abr17}.
Observe that the Gibbs state \eqref{eq:f_G} can be written in the form
\begin{equation}
f_G(\BV x_1,\BV v_1,\BV x_2,\BV v_2)=\frac{V^2}Ze^{-\phi(\|\BV x_2-\BV
  x_1\|)/ \theta_0}f_{1,G}(\BV v_1)f_{2,G}(\BV v_2),
\end{equation}
where $f_{1,G}$ and $f_{2,G}$ are the single-particle marginal densities,
\begin{equation}
f_{1,G}(\BV v)=f_{2,G}(\BV v)=\frac 1{(2\pi\theta_0)^{3/2}V}e^{-\|\BV
  v\|^2/2\theta_0},
\end{equation}
and $V$ is the volume of the coordinate domain. If the state $f$ is
close to the Gibbs equilibrium \eqref{eq:f_G}, we can assume that $f$
has the same marginal structure as \eqref{eq:f_G} above:
\begin{equation}
\label{eq:f_closure}
f(\BV x_1,\BV v_1,\BV x_2,\BV v_2)=\frac{V^2}Z e^{-\phi(\|\BV x_2 -\BV
  x_1\|)/ \theta_0} f_1(\BV x_1,\BV v_1)f_2(\BV x_2,\BV v_2),
\end{equation}
where $f_2$ is the marginal density of the second particle.
Substituting the closure for $f$ in \eqref{eq:f_closure} into
\eqref{eq:f_1} yields the Vlasov equation for the marginal density of
the first particle,
\begin{equation}
\label{eq:vlasov}
\parderiv{f_1}t+\BV v\cdot\parderiv{f_1}{\BV x_1}=\parderiv{\bar
  \phi(\BV x_1)}{\BV x_1}\cdot\parderiv{f_1}{\BV v_1},
\end{equation}
with $\bar\phi$ being the following mean field potential:
\begin{equation}
\label{eq:bphi}
\bar\phi(\BV x)=-\frac{V^2}{Z}\theta_0\int e^{-\phi(\|\BV x-\BV y\|)/
  \theta_0}\rho_2(\BV y)\dif\BV y,\qquad\rho_2(\BV y)=\int f_2(\BV y,
\BV w)\dif\BV w.
\end{equation}
While here the derivation of the Vlasov equation is presented for the
two-particle dynamics, in Appendix \ref{app:vlasov} we sketch the
general procedure for a multiparticle system.

The chief difference between $\phi$ and $\bar\phi$ is that, in the
convolution \eqref{eq:bphi} for the latter, the mass density $\rho_2$
acts as a low-pass filter. Thus, even if $\phi$ has typical properties
of an interaction potential (that is, peaked at zero, and a generally
rich Fourier spectrum), the mean field potential $\bar\phi$ is a large
scale potential, that is, it is confined to only a few large scale
Fourier modes.

At the same time, the form of the Vlasov equation \eqref{eq:vlasov} is
almost identical to that of the Liouville equation
\eqref{eq:liouville}, with the only difference that the interaction
potential $2\phi$ in the latter is replaced with $\bar\phi$ in the
former. While we note that, generally, $\bar\phi$ is time-dependent
(since it includes $\rho_2$), however, here it shall be assumed that,
on the relevant time scales, $\bar\phi$ can be regarded as constant in
time, and thus \eqref{eq:liouville} and \eqref{eq:vlasov} can be
treated in the same manner. So, while in what follows we refer chiefly
to the Liouville equation in \eqref{eq:liouville}, it also applies to
the Vlasov equation in \eqref{eq:vlasov}.

\section{The equation for the turbulent velocity}
\label{sec:turbulent_velocity}

In the conventional approach to fluid mechanics, the Boltzmann
equation is converted into a hierarchy of the transport equations for
the velocity moments, which is subsequently truncated at a suitable
point. Depending on where and how the moment hierarchy is truncated,
one obtains the Euler equations \citep{Bat}, the Navier--Stokes
equations \citep{Gols}, Grad's 13-moment system \citep{Gra}, the
regularization of Grad's 13-moment system \citep{StruTor}, etc. The
main tool in justifying such a truncation of the moment hierarchy is
Boltzmann's $H$-theorem \citep{Bol,Cer,CerIllPul}. Namely, in the
presence of the entropy growth, one argues that the distribution of
the molecular velocities in the Boltzmann equation must approach the
Maxwell--Boltzmann thermodynamic equilibrium state, and, therefore,
only a relatively few low-order velocity moments are sufficient to
accurately describe the overall shape of the solution.

In the present context, the role of the $H$-theorem is played by the
conservation of the R\'enyi metrics in \eqref{eq:renyi_conservation}.
Even though the solution of \eqref{eq:liouville} does not approach the
steady state \eqref{eq:f_0t} or \eqref{eq:f_Gt}, it also does not
escape it, maintaining, instead, a constant ``distance'' to the latter
in the sense of a R\'enyi metric. Clearly, if the initial condition
of~\eqref{eq:liouville} is close, in the sense of
\eqref{eq:renyi_conservation}, to \eqref{eq:f_0t} or \eqref{eq:f_Gt},
then the corresponding solution will also remain a nearby state. Thus,
in what follows, we apply the velocity moment procedure to
\eqref{eq:liouville} in a similar way it is done conventionally.

We define the velocity average $\langle a\rangle$ of a function $a(\BV
v)$ via
\begin{equation}
\langle a\rangle(t,\BV x)=\int a(\BV v)f(t,\BV x,\BV v)\dif\BV v.
\end{equation}
As usual, we denote the zero- and first-order velocity moments via the
density $\rho$ and average velocity $\BV u$, respectively:
\begin{equation}
\rho=\langle 1\rangle,\qquad\rho\BV u=\langle\BV v\rangle.
\end{equation}
Then, for the moments of $f$ in \eqref{eq:liouville} (or for those of
$f_1$ in \eqref{eq:vlasov}), we integrate the potential forcing terms
with $\BV v$-derivatives by parts, and obtain
\begin{subequations}
\label{eq:moment_equations}
\begin{equation}
\label{eq:mass_momentum}
\parderiv\rho t+\nabla\cdot(\rho\BV u)=0,\qquad\parderiv{(\rho\BV u)
}t+\nabla\cdot\big\langle\BV v^2\big\rangle=-2\rho\nabla\phi,
\end{equation}
\begin{equation}
\label{eq:energy}
\parderiv{\langle\BV v^2\rangle}t+\nabla\cdot\langle\BV v^3\rangle
=-2\rho\big(\BV u\nabla\phi^T+\nabla\phi\BV u^T\big),
\end{equation}
\end{subequations}
where $\BV v^2=\BV v\BV v^T$ is the outer product of $\BV v$ with
itself, and the symmetric 3-rank tensor $\BV v^3$ is the outer product
of $\BV v$ with itself, computed twice.

Next, let us decompose the quadratic moment $\langle\BV v^2\rangle$ as
\begin{equation}
\langle\BV v^2\rangle=\rho\BV u^2+\langle(\BV v-\BV u)^2\rangle,\quad
\text{with}\quad\langle(\BV v-\BV u)^2\rangle=\rho(\theta\BM I+\BM S).
\end{equation}
Above, $\theta$ is the kinetic temperature of $\BV v$, given via the
trace of the centered quadratic moment,
\begin{equation}
\rho\theta=\frac 13\trace\langle(\BV v-\BV u)^2\rangle=\frac 13
\langle\|\BV v-\BV u\|^2\rangle,
\end{equation}
while the matrix $\BM S$ is called the shear stress, and quantifies
the deviation of the centered quadratic moment from its own trace. By
construction, the trace of $\BM S$ is zero. The product $\rho\theta$
is known as the pressure.

Similar manipulations can be made for the cubic moment. Here, we
decompose
\begin{equation}
\langle\BV v^3\rangle=\rho\Big(\BV u^3+(\theta\BM I+\BM S)\otimes\BV
u+\big((\theta\BM I+\BM S)\otimes\BV u\big)^T+\big((\theta \BM I+\BM
S)\otimes\BV u\big)^{TT}\Big)+\langle(\BV v-\BV u)^3\rangle,
\end{equation}
where ``$T$'' and ``$TT$'' denote the two cyclic permutations of a
3-rank tensor, and ``$\otimes$'' denotes an outer product. In turn,
the skewness (that is, the centered cubic moment) $\langle(\BV v-\BV
u)^3\rangle$ can be expressed via
\begin{equation}
\langle(\BV v-\BV u)^3\rangle=\frac 25\rho\big(\BV q\otimes \BM I+(\BV
q\otimes\BM I)^T+(\BV q\otimes\BM I)^{TT}\big)+\rho\BM Q.
\end{equation}
Above, $\BV q$ is the conductive heat flux, given via
\begin{equation}
\rho\BV q=\frac 12\langle\|\BV v-\BV u\|^2(\BV v-\BV u) \rangle,
\end{equation}
while $\BM Q$ is the deviator between the heat flux and the skewness
tensor. Observe that $\BM Q$ is a symmetric 3-rank tensor whose
contraction along any pair of indices is zero. This is due to the fact
that the heat flux $\BV q$ is itself a multiple of the contracted
skewness moment:
\begin{equation}
(\rho\BV q)_i=\frac 12\sum_{j=1}^3\langle(\BV v-\BV u)^3
  \rangle_{jji}.
\end{equation}
The deviator $\BM Q$ is related to the centered cubic moment in the
same way as the shear stress $\BM S$ is related to the centered
quadratic moment -- namely, $\BM Q$ quantifies the deviation of the
skewness $\langle(\BV v-\BV u)^3\rangle$ from its own trace along any
pair of its indices.

\subsection{Nondimensionalization and scalings}

Following the standard approach in fluid mechanics, here we need to
choose suitable closures for $\BM S$ and $\BV q$. If the irreversible
collisions are present in the form of a Boltzmann collision integral,
one can assume that $\BM S$ and $\BV q$ are the steady states of their
respective transport equations \citep{Gols,Gra}.  Such a formalism
yields the Newton law of viscosity, and the Fourier law of heat
conduction. In our situation, however, there are no collision
integrals, and an analogous closure cannot be used. Instead, here we
close the velocity moment hierarchy in a novel way, making use of the
well-established criterion which holds systematically for observed
turbulent flows -- namely, the high Reynolds number condition.

Let $L$ and $U$ denote, respectively, the characteristic length scale
of the flow, and its reference speed, such that the ratio $U/L$
specifies the characteristic time scale. First, we rescale the time
and space differentiation operators, as well as the velocity, via
\begin{equation}
\nabla\to\frac 1L\nabla,\qquad\parderiv{}t\to\frac UL\parderiv{}t,
\qquad\BV u\to U\BV u.
\end{equation}
It is easy to see that the mass transport equation in
\eqref{eq:mass_momentum} is invariant with respect to this scaling.
Also, observe that it is pointless to rescale the density $\rho$ (as
the scaling constant would simply factor out of all transport
equations), and thus we leave it as is.

In the momentum equation, we need to choose the scalings for $\phi$,
$\theta$ and $\BM S$. For the potential $\phi$ and temperature
$\theta$, we choose the reference temperature constant $T$. The
treatment of the viscous shear stress $\BM S$ must be different, since
it quantifies an entirely different physical property -- namely, while
the temperature is related to the force which a fluid exerts onto a
plate in the orthogonal direction (the pressure), the shear stress
quantifies the tangential force of resistance to a shear (similar to
friction). Therefore, to rescale the stress $\BM S$, we introduce a
reference constant $\Nu$ which has the units of the kinematic
viscosity (that is, squared length over time), and choose the
following scaling:
\begin{equation}
\phi\to T\phi,\qquad\theta\to T\theta,\qquad\BM S\to\frac{\Nu U}L\BM
S.
\end{equation}
In the rescaled variables, the momentum equation in
\eqref{eq:mass_momentum} becomes
\begin{equation}
\label{eq:momentum_rescaled}
\parderiv{(\rho\BV u)}t+\nabla\cdot\bigg(\rho\BV u^2+\frac 1{\Mach^2}
\rho\theta\BM I+\frac 1\Rey\rho\BM S\bigg)=-\frac 2{\Mach^2}\rho\nabla
\phi,
\end{equation}
where $\Mach$ and $\Rey$ are the Mach and Reynolds numbers,
respectively:
\begin{equation}
\Mach=\frac U{\sqrt T},\qquad\Rey=\frac{UL}\Nu.
\end{equation}
Here note that our definition of the Mach number differs from the
traditional one by the square root of the adiabatic constant, for
convenience.  For the transport of higher-order moments, we need to
choose the scalings for $\BV q$ and $\BM Q$. In the absence of any
particular information about the difference in magnitudes between $\BV
q$ and $\BM Q$, we choose the rescaling for each quantity in the same
manner as above for $\theta$, with the additional multiplication by
the reference speed $U$ so that the physical units remain consistent:
\begin{equation}
\BV q\to TU\BV q,\qquad\BM Q\to TU\BM Q.
\end{equation}
Then, the transport equations for the pressure $\rho\theta$ and the
shear stress $\BM S$ become
\begin{subequations}
\begin{equation}
\label{eq:pressure_transport}
\parderiv{(\rho\theta)}t+\nabla\cdot(\rho\theta\BV u)+\frac 23\bigg[
  \rho\bigg(\theta\BM I+\frac{\Mach^2}\Rey\BM S\bigg):\nabla\BV u+
  \nabla\cdot(\rho\BV q)\bigg]=0,
\end{equation}
\begin{multline}
\label{eq:stress_transport}
\parderiv{(\rho\BM S)}t+\nabla\cdot(\BV u\otimes\BM S)+\rho\bigg(\BM S
\nabla\BV u+\nabla\BV u^T\BM S-\frac 23(\BM S:\nabla\BV u)\BM I
\bigg)+\frac\Rey{\Mach^2}\bigg[\nabla\cdot(\rho\BM Q)+\\+\frac 25
  \bigg(\nabla(\rho\BV q)+ \nabla(\rho\BV q)^T-\frac 23\nabla\cdot
  (\rho\BV q)\BM I\bigg)+\rho\theta\bigg(\nabla\BV u+\nabla\BV u^T
  -\frac 23(\nabla\cdot\BV u)\BM I\bigg)\bigg]=\BM 0,
\end{multline}
\end{subequations}
where ``$:$'' denotes the Frobenius product of two matrices.  The
equations above are obtained by subtracting appropriate combinations
of the momentum and density equations in \eqref{eq:mass_momentum} from
the transport equation for the quadratic moment \eqref{eq:energy} (for
more details, see \citet{Gra,Abr13,Abr17}, or \citet{AbrOtt}).

The chief difference between \eqref{eq:stress_transport} and the
stress transport equation originating from the Boltzmann equation
(see, for example, \citet{Gra}) is that the latter contains an
additional viscous damping term due to the time-irreversible effects
from the Boltzmann collision integral. The presence of such damping
term leads to the Newton law of viscosity in the Navier--Stokes
equations; here, however, we will need a different closure.

\subsection{Approximate relations for a high Reynolds number}

It is known from observations that the turbulence manifests itself at
high Reynolds numbers, $\Rey>10^3$ \citep{Rey}. Assuming that the
magnitude of all rescaled variables is of order one, and the Reynolds
number is high, we can infer the following approximate relations.
First, in the pressure transport equation
\eqref{eq:pressure_transport}, the term with the stress is divided by
$\Rey$, and thus should be much smaller than the rest of the terms.
Second, in the stress transport equation \eqref{eq:stress_transport},
the group of terms which is multiplied by $\Rey$ would normally be
much larger than the rest of the terms, which would cause the time
derivative of the stress to be large, which, in turn, would likely
result in $\BM S$ itself to grow large. Such growth would mean that
the flow became viscous, rather than turbulent. Conversely, the
requirement that the flow is turbulent suggests that the terms
multiplied by $\Rey$ should add up to zero, which, together with the
fact that the trace of $\BM Q$ along any pair of its indices is zero,
leads to the following relations:
\begin{subequations}
\label{eq:heat_flux_closure}
\begin{equation}
\frac 25\big(\nabla(\rho\BV q)+\nabla(\rho\BV
q)^T\big)+\nabla\cdot(\rho\BM Q)=-\rho\theta\big(\nabla\BV
u+(\nabla\BV u)^T\big),
\end{equation}
\begin{equation}
\label{eq:qdiv_closure}
\frac 25\nabla\cdot(\rho\BV q)=-\rho\theta\nabla\cdot\BV u.
\end{equation}
\end{subequations}
It turns out that, in our model, the high Reynolds number relations in
\eqref{eq:heat_flux_closure} fully define the closure for the heat
flux, and lead to the transport equation for the turbulent velocity
alone. First, the substitution of \eqref{eq:qdiv_closure} into the
pressure transport equation \eqref{eq:pressure_transport}, and the
removal of the stress term (which is divided by $\Rey$) in the latter,
lead to
\begin{equation}
\label{eq:pressure}
\parderiv{(\rho\theta)}t+\BV u\cdot\nabla(\rho\theta)=0,\qquad
\parderiv{(\theta^{-1})}t+\nabla\cdot\big(\theta^{-1}\BV u\big)=0.
\end{equation}
Clearly, the equation for the pressure $\rho\theta$ above indicates
that the pressure is constant along the streamlines, which, in turn,
means that the gradient of the pressure at a given point is always
orthogonal to the corresponding streamline which passes through the
same point. Next, let us write the momentum equation in
\eqref{eq:mass_momentum} in the form
\begin{equation}
\label{eq:u}
\parderiv{\BV u}t+\BV u\cdot\nabla\BV u+\frac 1{\Mach^2\rho}
\big(\nabla(\rho\theta)+2\rho\nabla\phi\big)=\BV 0,
\end{equation}
where we used the mass transport equation to eliminate the time
derivative of $\rho$. As the pressure gradient is always orthogonal to
the direction of the flow, and thus has no effect on it, we conclude
that the flow must be hydrostatically balanced, separating the above
equation into two as follows:
\begin{equation}
\label{eq:hydrostatic_balance}
\parderiv{\BV u}t+\BV u\cdot\nabla\BV u=\BV 0,\qquad
\nabla(\rho\theta)+2\rho\nabla\phi=\BV 0.
\end{equation}
Observe, however, that the approximate relations in
\eqref{eq:heat_flux_closure}--\eqref{eq:hydrostatic_balance}, which
are based on the assumption of a high Reynolds number, render the
relevant transport equations to be unrealistic. For example, even
though the equation for the velocity in \eqref{eq:hydrostatic_balance}
is completely decoupled from all other variables, its solutions must
nonetheless be tangent to the level sets of $\phi$. Moreover, the
density $\rho$ and the temperature $\theta$ effectively become
independent variables, transported by $\BV u$ (in fact, the equations
for $\theta^{-1}$ in \eqref{eq:pressure} and for $\rho$ in
\eqref{eq:mass_momentum} are identical), yet, they still have to be
connected via the hydrostatic balance relation
\eqref{eq:hydrostatic_balance}, which does not even include $\BV
u$. This situation happens because, by implementing the relations in
\eqref{eq:heat_flux_closure}--\eqref{eq:hydrostatic_balance} directly,
we place the restrictions onto the time-derivative of the stress $\BM
S$ in its transport equation \eqref{eq:stress_transport}, rather than
onto $\BM S$ itself.

\subsection{The closure for the second time-derivative}

Technically, if the magnitude of $\BM S$ itself must be controlled,
the appropriate restrictions must be placed on the time integral of
the multiple of $\Rey$ in \eqref{eq:stress_transport}, which is a much
weaker constraint. The time-derivative of $\BM S$ can be permitted to
have rapid oscillations around zero, so long as they are canceled out
by the time integration. As a direct consequence, same considerations
apply to the heat flux \eqref{eq:heat_flux_closure}, pressure
transport \eqref{eq:pressure} and hydrostatic balance
\eqref{eq:hydrostatic_balance} relations.

However, placing the constraints on the time integrals of given
quantities is practically difficult, if the goal is to derive a PDE,
rather than a more general integro-differential equation.  Here,
instead, we will take a simpler approach, by implementing the same
constraints as above in \eqref{eq:heat_flux_closure} and
\eqref{eq:hydrostatic_balance} in the equation for the second
time-derivative of the velocity instead; this is somewhat analogous to
restricting their time integrals in a first-order equation. Observe
that, above in \eqref{eq:u}--\eqref{eq:hydrostatic_balance}, the
restrictions are imposed onto the equation of the form
\begin{equation}
\parderiv{\BV u}t + \text{advection} = \text{forcing},
\end{equation}
that is, the constraints are placed on the rate of change of the
velocity $\BV u$ directly. Below, instead, we will implement the
restrictions in the equation of the form
\begin{equation}
\label{eq:u_advection_forcing}
\parderiv{^2\BV u}{t^2} + \text{advection} = \text{forcing},
\end{equation}
which itself is obtained via the time-differentiation of the
dynamics. The main difference between these two equations is that the
latter can, generally, have secular (that is, slowly,
sub-exponentially growing) terms, which do not develop in the
former. Below we will observe that the turbulent spectra seem to
manifest precisely in such secular dynamics.

In the second order equation for the velocity, we will implement the
following two constraints: the heat flux closure in
\eqref{eq:heat_flux_closure}, and the hydrostatic balance relation for
the pressure gradient in \eqref{eq:hydrostatic_balance}. We will,
however, not assume that $\BV u$ is necessarily constrained to the
level sets of $\phi$, only that the hydrostatic balance between the
pressure gradient and the potential forcing holds irrespectively of
the direction of the flow.

To derive the suitable second-order equation, we start by
time-differentiating the momentum equation in
\eqref{eq:momentum_rescaled}:
\begin{equation}
\parderiv{^2(\rho\BV u)}{t^2}+\parderiv{}t\nabla\cdot\bigg(\rho\BV
u^2+\frac{\rho\theta}{\Mach^2}+\frac{\rho\BM S}\Rey\bigg)=\frac
2{\Mach^2}\nabla \cdot(\rho\BV u)\nabla\phi.
\end{equation}
Above, we used the density equation to replace the time derivative of
the density in the right-hand side with the negative divergence of the
momentum. To express the time derivative of the divergence of the
quadratic moment, we compute the divergence of~\eqref{eq:energy}:
\begin{multline}
\parderiv{}t\nabla\cdot\bigg(\rho\BV u^2+\frac{\rho\theta}{\Mach^2}+
\frac{\rho\BM S}\Rey\bigg)=-\frac 2{\Mach^2}\nabla\cdot\big[\rho\big(
  \BV u(\nabla\phi )^T+(\nabla \phi)\BV u^T\big)\big]-\nabla^2:(\rho
\BV u^3)-\\-\frac 1\Rey\nabla^2:\Big[\rho\Big(\BM S\otimes\BV u+(\BM S
  \otimes\BV u)^T+(\BM S\otimes\BV u)^{TT}\Big)\Big]-\frac 1{\Mach^2}
\nabla\cdot\bigg\{\nabla\bigg[\rho\bigg(\theta\BV u+\frac 25\BV q
  \bigg)\bigg]+\\+\nabla\bigg[\rho\bigg(\theta\BV u+\frac 25\BV q
  \bigg)\bigg]^T+\nabla \cdot\bigg[\rho\bigg(\theta\BV u+\frac 25\BV
  q\bigg)\bigg]\BM I+ \nabla\cdot(\rho\BM Q)\bigg\}.
\end{multline}
Combining the above two equations via the mixed derivative of the
quadratic moment, we arrive at
\begin{multline}
\label{eq:momentum_2}
\parderiv{^2(\rho\BV u)}{t^2}-\nabla^2:(\rho \BV u^3)-\frac
1\Rey\nabla^2:\Big[\rho\Big(\BM S\otimes\BV u+(\BM S \otimes\BV
  u)^T+(\BM S\otimes\BV u)^{TT}\Big)\Big]-\\-\frac 1{\Mach^2}
\nabla\cdot\bigg\{\nabla\bigg[\rho\bigg(\theta\BV u+\frac 25\BV q
  \bigg)\bigg]+\nabla\bigg[\rho\bigg(\theta\BV u+\frac 25\BV q
  \bigg)\bigg]^T+\nabla \cdot\bigg[\rho\bigg(\theta\BV u+\frac 25\BV
  q\bigg)\bigg]\BM I+ \nabla\cdot(\rho\BM Q)\bigg\}=\\=\frac
2{\Mach^2}\nabla \cdot(\rho\BV u)\nabla\phi+\frac
2{\Mach^2}\nabla\cdot\big[\rho\big( \BV u(\nabla\phi )^T+(\nabla
  \phi)\BV u^T\big)\big].
\end{multline}
Next, we expand the second time-derivative of the momentum as
\begin{equation}
\parderiv{^2(\rho\BV u)}{t^2}=\rho\parderiv{^2\BV u}{t^2}+2\parderiv
\rho t\parderiv{\BV u}t+\parderiv{^2\rho}{t^2}\BV u,
\end{equation}
and eliminate the first time-derivative of the velocity and the second
time-derivative of the density via the transport equations for the
mass and momentum:
\begin{subequations}
\label{eq:momentum_22}
\begin{equation}
\parderiv{^2\rho}{t^2}=-\nabla\cdot\bigg(\parderiv{(\rho\BV u)}t\bigg)
=\nabla^2:(\rho\BV u^2)+\nabla\cdot\bigg(\frac{\nabla(\rho\theta)}
{\Mach^2}+\frac{\nabla\cdot(\rho\BM S)}\Rey+\frac{2\rho\nabla\phi}{
  \Mach^2}\bigg),
\end{equation}
\begin{equation}
\parderiv{\BV u}t=\frac 1\rho\left( \parderiv{(\rho\BV u)}t-\parderiv
\rho t\BV u\right)=-\BV u\cdot\nabla\BV u-\frac 1\rho \bigg(\frac{
  \nabla(\rho\theta)}{\Mach^2}+\frac{\nabla\cdot(\rho\BM S)}\Rey+
\frac{2\rho\nabla\phi}{\Mach^2} \bigg).
\end{equation}
\end{subequations}
Combining together \eqref{eq:momentum_2}--\eqref{eq:momentum_22}, and
observing that 
\begin{equation}
\nabla^2:(\rho\BV u^2)\BV u-\nabla^2:(\rho\BV u^3)+2\nabla\cdot(\rho
\BV u)(\nabla\BV u)\BV u=-\rho\BV u\cdot\big[(\nabla\BV u)^2+\nabla
  (\BV u\cdot\nabla \BV u)\big],
\end{equation}
we arrive at
\begin{multline}
\label{eq:u2}
\parderiv{^2\BV u}{t^2}-\BV u\cdot\big[(\nabla\BV u)^2+\nabla(\BV u
  \cdot\nabla \BV u)\big]+\frac 2{\rho^2}\nabla\cdot(\rho\BV u)\bigg(
\frac{\nabla(\rho\theta)}{ \Mach^2}+\frac{\nabla\cdot(\rho\BM S)}\Rey
+\frac{2\rho\nabla\phi}{\Mach^2}\bigg)+\\+\nabla\cdot\bigg(\frac{\nabla
  (\rho\theta)}{\Mach^2}+\frac{\nabla\cdot(\rho\BM S)}\Rey+\frac{2\rho
  \nabla\phi}{\Mach^2}\bigg)\frac{\BV u}\rho-\frac 1{\Rey\rho}\nabla^2
:\Big[\rho\Big(\BM S\otimes\BV u+(\BM S\otimes\BV u)^T+(\BM S\otimes
  \BV u)^{TT}\Big)\Big]-\\-\frac 1{\Mach^2\rho}\nabla\cdot\bigg\{
\nabla\bigg[\rho\bigg(\theta\BV u+\frac 25\BV q\bigg)\bigg]+\nabla
\bigg[\rho\bigg(\theta\BV u+\frac 25\BV q \bigg)\bigg]^T+\nabla\cdot
\bigg[\rho\bigg(\theta\BV u+\frac 25\BV q\bigg)\bigg]\BM I+\nabla
\cdot(\rho\BM Q)\bigg\}=\\=\frac 2{\Mach^2\rho}\big\{\nabla\cdot(\rho
\BV u)\nabla\phi+\nabla\cdot\big[\rho\big(\BV u(\nabla\phi)^T+(\nabla
  \phi)\BV u^T\big)\big]\big\}.
\end{multline}
Observe that the first two terms above constitute the pure velocity
advection from the hydrostatic balance equation
\eqref{eq:hydrostatic_balance}, differentiated in time; indeed,
\begin{equation}
\label{eq:u_no_forcing}
\parderiv{}t\bigg(\parderiv{\BV u}t+\BV u\cdot\nabla\BV
u\bigg)=\parderiv{^2\BV u}{t^2}+\parderiv{}t(\BV u\cdot \nabla\BV
u)=\parderiv{^2\BV u}{t^2}-\BV u\cdot\big[(\nabla\BV u)^2 +\nabla(\BV
  u\cdot\nabla \BV u)\big],
\end{equation}
where \eqref{eq:hydrostatic_balance} was used in the last identity.
Thus, the form of \eqref{eq:u2} conforms to
\eqref{eq:u_advection_forcing}, with the forcing comprised of the rest
of the terms involving $\rho$, $\theta$, $\BM S$, $\BV q$, $\BM Q$ and
$\phi$.

The next step is to simplify the forcing term in \eqref{eq:u2}. For
that, we first remove the terms divided by the Reynolds number, and
use the heat flux closure \eqref{eq:heat_flux_closure}:
\begin{multline}
\parderiv{^2\BV u}{t^2}-\BV u\cdot\big[(\nabla\BV u)^2+\nabla(\BV u
  \cdot \nabla\BV u)\big]+\frac 2{\Mach^2\rho^2}\nabla\cdot(\rho\BV u)
\big(\nabla(\rho\theta)+2\rho\nabla\phi\big)+\\+\frac 1{\Mach^2\rho}
\nabla\cdot\big(\nabla(\rho\theta)+2\rho\nabla\phi\big)\BV u-\frac 1{
  \Mach^2\rho}\nabla\cdot\big[\big(\nabla(\rho\theta)+2\rho\nabla\phi
  \big)\BV u^T+\BV u\big(\nabla(\rho\theta)+2\rho\nabla\phi\big)^T
  \big]=\\=\frac 1{\Mach^2\rho}\big[\nabla\big(\BV u\cdot\nabla(\rho
  \theta)\big)+2\nabla\cdot(\rho\BV u)\nabla\phi\big].
\end{multline}
Next, we use the hydrostatic balance relation in
\eqref{eq:hydrostatic_balance} to replace $\nabla(\rho\theta)$ with
$-2\rho\nabla\phi$ throughout the equation, which simplifies the
latter to
\begin{equation}
\label{eq:turbulent_velocity_2}
\parderiv{^2\BV u}{t^2}-\BV u\cdot\big[(\nabla\BV u)^2+\nabla(\BV u
  \cdot\nabla\BV u)\big]=\frac 2{\Mach^2}\left[-\BV u\cdot\nabla^2\phi
  +\frac 1\rho\big(\nabla\cdot(\rho\BV u)\BM I-\nabla(\rho\BV u)\big)
  \nabla\phi\right].
\end{equation}
In the last term above, we express $\nabla\rho/\rho$ via
\eqref{eq:hydrostatic_balance}, which leads to the identity
\begin{multline}
\label{eq:rho_theta_term}
\frac 1\rho\big(\nabla\cdot(\rho\BV u)\BM I-\nabla(\rho\BV u)\big)
\nabla\phi=(\nabla\cdot\BV u)\nabla\phi-\nabla\BV u\nabla\phi +\frac
1\rho\big(\nabla\phi\nabla\rho^T-\nabla\rho\nabla\phi^T\big)\BV u=\\=
(\nabla\cdot\BV u)\nabla\phi-\nabla\BV u\nabla\phi-\frac 1\theta\big(
\nabla\phi\nabla\theta^T-\nabla\theta\nabla\phi^T\big)\BV u=\theta
\big(\nabla\cdot(\theta^{-1}\BV u)\BM I-\nabla(\theta^{-1}\BV u)\big)
\nabla\phi.
\end{multline}
Here, $\nabla\cdot(\rho\BV u)$ and $\nabla\cdot(\theta^{ -1}\BV u)$
are the advection terms for $\rho$ in \eqref{eq:mass_momentum} and
$\theta^{-1}$ in \eqref{eq:pressure}, and thus their contribution
tends to average out to zero over time; this, in turn, means that
$\rho^{-1}\nabla(\rho\BV u)\nabla\phi\approx
\theta\nabla(\theta^{-1}\BV u)\nabla\phi$. Here, $\theta^{-1}$ is not
necessarily a multiple of $\rho$ (as follows, for example, from the
Gibbs equilibrium state), and, generally, the matrices
$\rho^{-1}\nabla(\rho\BV u)$ and $\theta\nabla(\theta^{-1}\BV u)$ are
expected to behave in an independent manner, as $\rho$ and
$\theta^{-1}$ are propagated by \eqref{eq:mass_momentum} and
\eqref{eq:pressure} independently. We thus assume that, for the
relation above to hold, $\nabla\phi$ must on average align with the
null spaces of both matrices. This brings the overall contribution
from \eqref{eq:rho_theta_term} to zero and leads to the following
equation for $\BV u$ alone:
\begin{equation}
\label{eq:turbulent_velocity}
\parderiv{^2\BV u}{t^2}-\BV u\cdot\big[(\nabla\BV u)^2+\nabla(\BV u
  \cdot\nabla\BV u)\big]=-\frac 2{\Mach^2}\BV u\cdot\nabla^2\phi.
\end{equation}
The fact that we arrived at a standalone equation for the turbulent
velocity, independent of the other physical parameters, is consistent
with the fact that the observed turbulence appears to have largely
universal properties across a variety of different media.

The turbulent velocity equation \eqref{eq:turbulent_velocity} does not
have any apparent dissipative terms. For the purpose of the current
work, we avoid introducing any {\em ad hoc} form of damping into
\eqref{eq:turbulent_velocity} solely to stabilize its solution, since
any dissipation effect must originate systematically from the
underlying dynamics in \eqref{eq:dyn_sys}. Instead, we have to keep in
mind that the time scale, on which the solutions of
\eqref{eq:turbulent_velocity} are physically meaningful, is limited.

\section{Numerical simulation of the turbulent velocity equation}
\label{sec:numerical}

Here we show the results of several numerical simulations of the
turbulent velocity equation \eqref{eq:turbulent_velocity} with
different types of the interaction potential in a periodic unit
cube. In~\eqref{eq:turbulent_velocity}, observe that the Mach number
only appears in the potential forcing term. Since below we will choose
the magnitude of the potential $\phi$ artificially to put
\eqref{eq:turbulent_velocity} into a desired dynamical regime, we set
the Mach number to unity, $\Mach=1$.

The initial condition, chosen for the velocity field in all cases
below, is a periodic laminar shear flow, with the fluid moving in the
$y$-direction, and varying in the $x$-direction:
\begin{equation}
\label{eq:u_sinx}
\BV u_{t=0}=(0,\sin(2\pi x),0).
\end{equation}
The choice of the shear function is based on the following two
requirements: first, it must be periodic, and, second, it must decay
to zero at $\BV x=\BV 0$, since, in the present context, $\BV u$ is
the average difference between the particle velocities as a function
of the distance, and thus we expect it to vanish when the distance
itself vanishes.  Arguably, the simplest large-scale function which
meets these two criteria is the sine.

For the numerical simulation, we partition the domain into 150 uniform
discretization steps in each of the three directions, which results in
$150^3=3375000$ discretization cubes. To carry out the numerical
simulations, we use OpenFOAM \citep{WelTabJasFur}, which provides a
variety of finite volume methods to discretize large sets of data, as
well as facilities for a convenient parallelization of computations.
The forward time integration of \eqref{eq:turbulent_velocity} in all
cases below is carried out by the standard OpenFOAM built-in scheme
for the second time derivative with the fixed time step of $\Delta
t=0.01$ units.

\subsection{Electrostatic potential}

The electrostatic potential is given via $\phi(r)=\phi_0/r$, however,
it has a singularity at zero. For the purposes of the numerical
simulation, we ``regularize'' $\phi(r)$ near zero as follows:
\begin{equation}
\label{eq:reg_potential}
\phi(r)=\left\{\begin{array}{l@{\qquad}l} \displaystyle
\frac{\phi_0}r, & r\geq r_0, \\ \displaystyle \frac
12\frac{\phi_0}{r_0} \bigg( 3-\frac{r^2}{r_0^2}\bigg), & 0\leq
r<r_0.\end{array}\right.
\end{equation}
As we can see, the potential $\phi(r)$ behaves as $\sim r^{-1}$ as
long as $r\geq r_0$, and is capped by an inverted parabola for $0\leq
r<r_0$ to avoid the singularity at zero. The parameters of the
parabola are chosen to match the value and the first derivative of
$\phi_0/r$ at $r_0$.

The constants $\phi_0$ and $r_0$ are chosen as follows:
\begin{equation}
\phi_0=10^{-8},\qquad r_0=5\cdot 10^{-3}.
\end{equation}
Observe that $r_0$ defines the length scale beyond which the Fourier
transform of the potential start decaying rapidly. Since the minimal
scale in our model, due to the finite discretization, is $1/150\sim
7\cdot 10^{-3}$, the choice of $r_0$ ensures that the decay of the
Fourier transform of $\phi(r)$ corresponds to that of the
electrostatic potential in the whole range of the length scales of the
model.

We, however, found that, in our finite periodic domain, the Fourier
transform of the regularized potential \eqref{eq:reg_potential}
exhibits small, but rapid oscillations throughout the whole range of
its Fourier coefficients due to discontinuity in the first derivative
at the edges of the periodic domain. To remove those oscillations and
make the behavior of the Fourier transform of our potential more
similar to that of the pure electrostatic potential in an unbounded
domain, in the numerical model we further adjusted the regularized
potential $\phi(r)$ from \eqref{eq:reg_potential} as follows:
\begin{equation}
\label{eq:num_potential}
\phi_{num}(r)=\left\{\begin{array}{l@{\qquad}l} \phi(r)-c, &
\phi(r)-c\geq 0, \\ 0, & \phi(r)-c<0,\end{array}
\right.\qquad\text{for }c=2\cdot 10^{-8}.
\end{equation}
As we can see, the numerical model potential in
\eqref{eq:num_potential} is slightly lower than the unadjusted one in
\eqref{eq:reg_potential}, and is set to zero whenever the lowered
value dips below zero (which only happens at the very edges of the
domain).

\begin{figure}
\includegraphics[width=\textwidth]{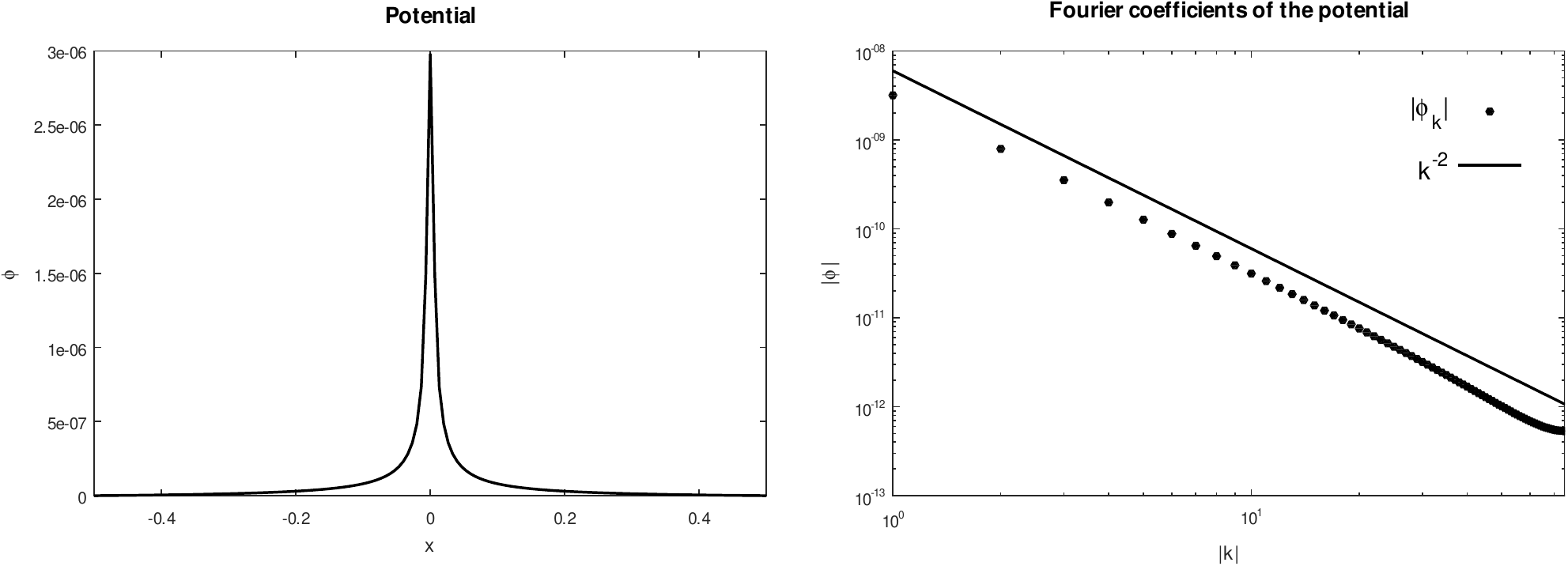}
\caption{The graph of the regularized electrostatic potential
  $\phi_{num}(|x|)$ (left), and its Fourier transform
  $\hat\phi_{num}(\|\BV k\|)$ (right). The line $\|\BV k\|^{-2}$
  is given for reference.}
\label{fig:electrostatic_potential}
\end{figure}

For the chosen values of $\phi_0$, $r_0$ and $c$, the graph of the
model potential $\phi_{num}(r)$ and its Fourier transform
$\hat\phi_{num}(\|\BV k\|)$ are shown in Figure
\ref{fig:electrostatic_potential}.  Observe that the decay of the
Fourier transform $\hat\phi_{num}(\|\BV k\|)$ corresponds to $\|\BV
k\|^{-2}$ without any visible oscillations, and is similar to the
decay of the pure electrostatic potential in an unbounded
domain~\citep{Abr20}.

\begin{figure}
\includegraphics[width=0.5\textwidth]{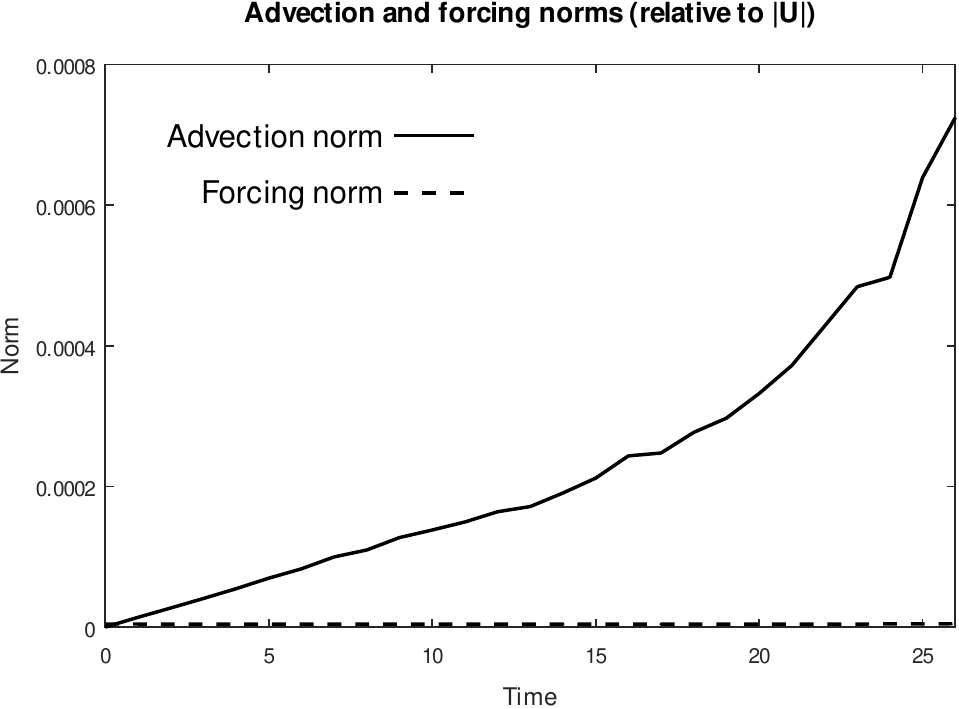}
\caption{The time series of the quadratic norms of the advection part
  of~\eqref{eq:turbulent_velocity} (solid line), and its potential
  forcing part (dashed line), for the electrostatic potential. Both
  norms are given relative to the norm of the turbulent velocity $\BV
  u$ itself.}
\label{fig:electrostatic_norms}
\end{figure}
\begin{figure}
\includegraphics[width=\textwidth]{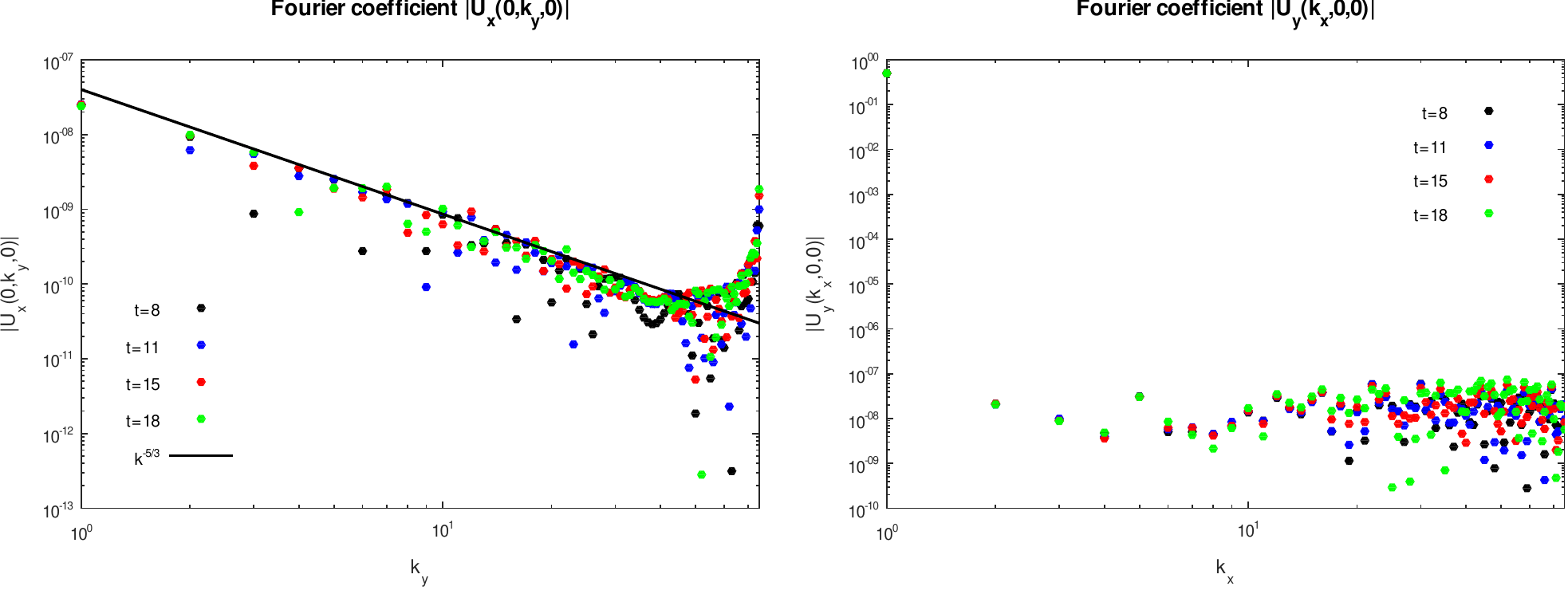}
\caption{The values of $|\hat u_x(0,k_y,0)|$ and $|\hat u_y(k_x,0,0)|$
  for the electrostatic potential forcing, captured at times
  $t=8,11,15,18$. The line $k_y^{-5/3}$ is given for the
  reference.}
\label{fig:electrostatic_velocity}
\end{figure}
\begin{figure}
\includegraphics[width=\textwidth]{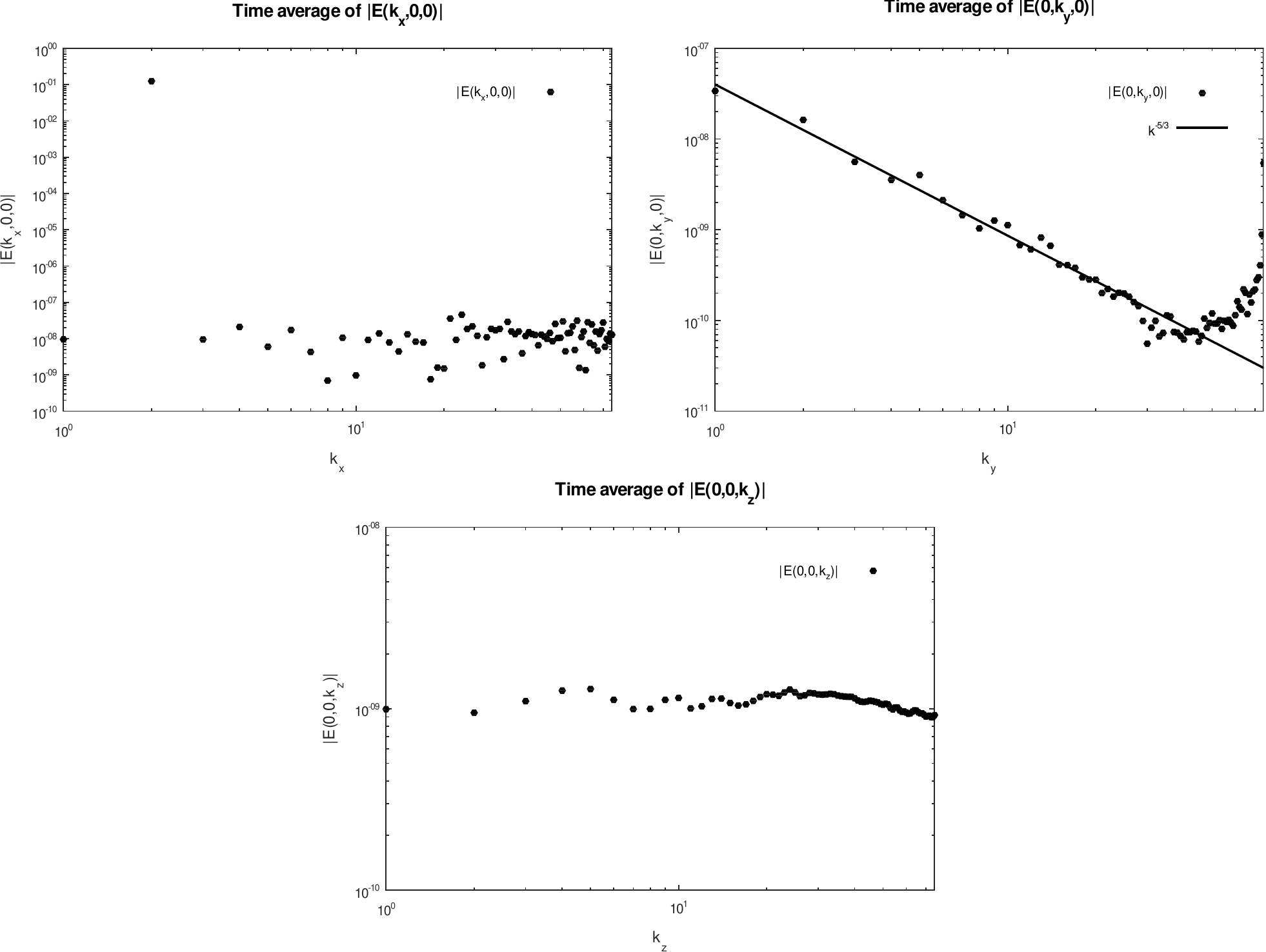}
\caption{The values of $|\hat E(k_x,0,0)|$, $|\hat E(0,k_y,0)|$ and
  $|\hat E(0,0,k_z)|$ for the electrostatic potential forcing,
  averaged between the times $t=8$ and $t=18$. The line
  $k_y^{-5/3}$ is given for the reference.}
\label{fig:electrostatic_energy}
\end{figure}

As mentioned above, in the absence of the potential forcing, the
turbulent velocity equation in \eqref{eq:turbulent_velocity} is the
decoupled (under the hydrostatic balance relation
\eqref{eq:hydrostatic_balance}) advection, differentiated once more in
time \eqref{eq:u_no_forcing}. Thus, the laminar shear flow
\eqref{eq:u_sinx} would obviously be a steady state in such a
situation, so we set the initial condition for the first
time-derivative of the velocity to zero. This initial condition models
the scenario where a potential forcing is introduced to an otherwise
steady laminar shear flow.

Our first goal here is to locate the secular growth regime, which, in
the absence of any sort of damping, should preclude the exponential
growth and the resulting numerical blow-up. In order to do that, first
we find that our numerical simulation blows up shortly after time
$t=26$ units. To locate the regime with the secular growth, in
Figure~\ref{fig:electrostatic_norms} we plot the quadratic norms of
the advection and forcing terms in \eqref{eq:turbulent_velocity},
relative to the norm of the velocity itself, as functions of time.
Observe that, while the potential forcing norm remains close to a very
small constant, the advection norm grows linearly with time until
roughly 18-20 time units, after which a rapid, exponential growth
starts. Thus, the secular growth regime in this scenario extends
roughly until the time $t=18$ units.

Before proceeding to study the time averages of the Fourier transforms
of the kinetic energy, we examine the snapshots of the Fourier
transforms of the turbulent velocity. Now, observe that since, for a
given time value, $\BV u(\BV x)$ is a vector field, its Fourier
transform $\hat{\BV u}(\BV k)$ is also a vector field -- namely, not
only $\hat{\BV u}(\BV k)$ consists of the three scalar components
$\hat u_x(\BV k)$, $\hat u_y(\BV k)$ and $\hat u_z(\BV k)$, but also
each of these three components is a function of the three-dimensional
wavevector $\BV k$. As a result, $\hat{\BV u}(\BV k)$ may potentially
have a complex structure in the full three-dimensional space, which
would lead to an exhaustive study.

Thus, for the convenience of the reader, in the current work we
restrict the presented results only to those which would be relatively
easy to observe experimentally. In the literature
\citep{BucVel,NasGag}, the Fourier transforms are typically presented
along a given direction, which suggests that the observed data are
averaged over the other two directions, thus corresponding to zero
Fourier wavenumbers.  Thus, here we focus on the Fourier transforms
$\hat{\BV u}(k_x,0,0)$, $\hat{\BV u}(0,k_y,0)$ and $\hat{\BV
  u}(0,0,k_z)$, which are three-dimensional vector functions of a
scalar argument. Further, we find that the majority of the components
of these vector functions have typical values of the machine round-off
errors, and thus appear to belong to the null space of the dynamics,
for the given initial condition and the potential. Only the two
components, $\hat u_x(0,k_y,0)$ and $\hat u_y(k_x,0,0)$, are
discernibly nonzero.

In Figure \ref{fig:electrostatic_velocity}, we show the snapshots of
$|\hat u_x(0,k_y,0)|$ and $|\hat u_y(k_x,0,0)|$ at $t=8,11,15,18$,
which belong to the regime of the secular growth. As we can see,
although the component $|\hat u_x(0,k_y,0)|$ shows noticeable
scattering, the top of the scatterplot visibly decays as $k_y^{-5/3}$,
which corresponds to the Kolmogorov spectrum. On the other hand, the
component $\hat u_y(k_x,0,0)$ exhibits a largely flat spectrum, with
the exception of a single large scale value which corresponds to the
shear flow \eqref{eq:u_sinx}.

The kinetic energy of the flow $E(\BV x)=\|\BV u(\BV x)\|^2/2$, as
well as its Fourier transform $\hat E(\BV k)$, are scalar functions of
vector arguments, and thus it is relatively easy to examine the
structure of the latter along the axes. In Figure
\ref{fig:electrostatic_energy} we present the time averages of $|\hat
E(k_x,0,0)|$, $|\hat E(0,k_y,0)|$ and $|\hat E(0,0,k_z)|$, computed
between the times $t=8$ and $t=18$. Observe that, while the time
averages of $|\hat E(k_x,0,0)|$ and $|\hat E(0,0,k_z)|$ have largely
flat spectra, the time-average of $|\hat E(0,k_y,0)|$ aligns very well
with the reference line $k_y^{-5/3}$, which matches the Kolmogorov
spectrum. This result agrees with the experiment of \citet{BucVel},
who also observed the Kolmogorov spectrum of the kinetic energy along
the direction of the flow.

\subsection{Gravity potential}

\begin{figure}
\includegraphics[width=0.5\textwidth]{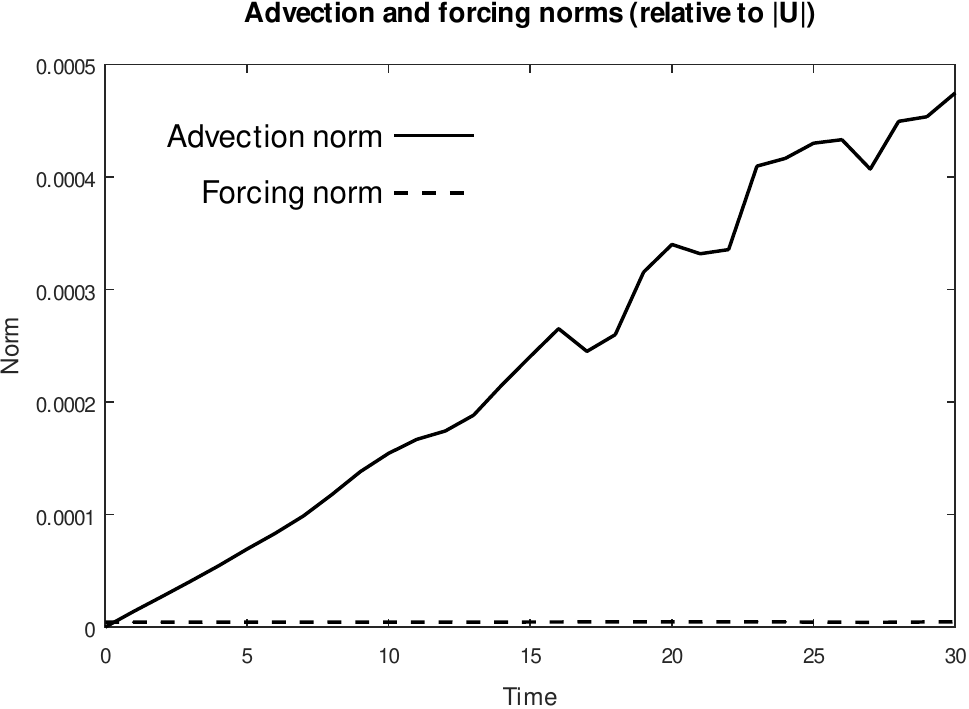}
\caption{The time series of the quadratic norms of the advection part
  of~\eqref{eq:turbulent_velocity} (solid line), and its potential
  forcing part (dashed line), for the gravity potential. Both norms
  are given relative to the norm of the turbulent velocity $\BV u$
  itself.}
\label{fig:gravity_norms}
\end{figure}
\begin{figure}
\includegraphics[width=\textwidth]{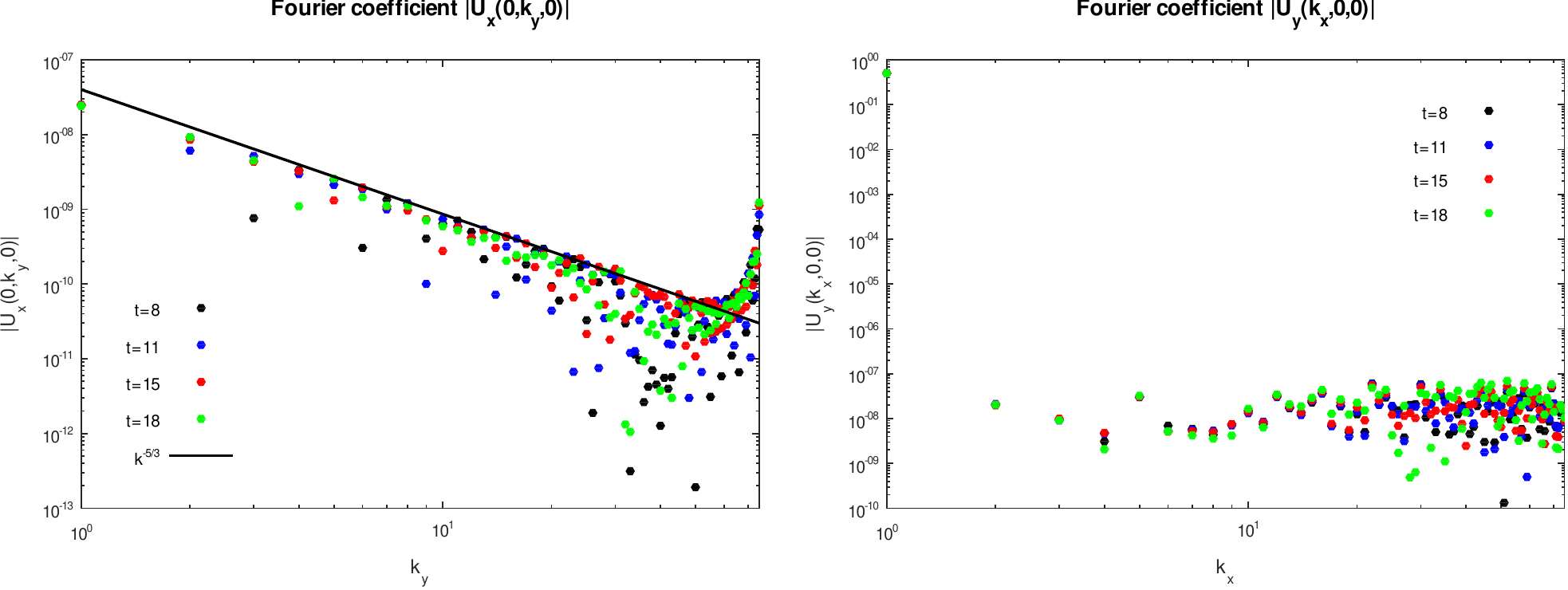}
\caption{The values of $|\hat u_x(0,k_y,0)|$ and $|\hat u_y(k_x,0,0)|$
  for the gravity potential forcing, captured at times $t=8,11,15,18$.
  The line $k_y^{-5/3}$ is given for the reference.}
\label{fig:gravity_velocity}
\end{figure}
\begin{figure}
\includegraphics[width=\textwidth]{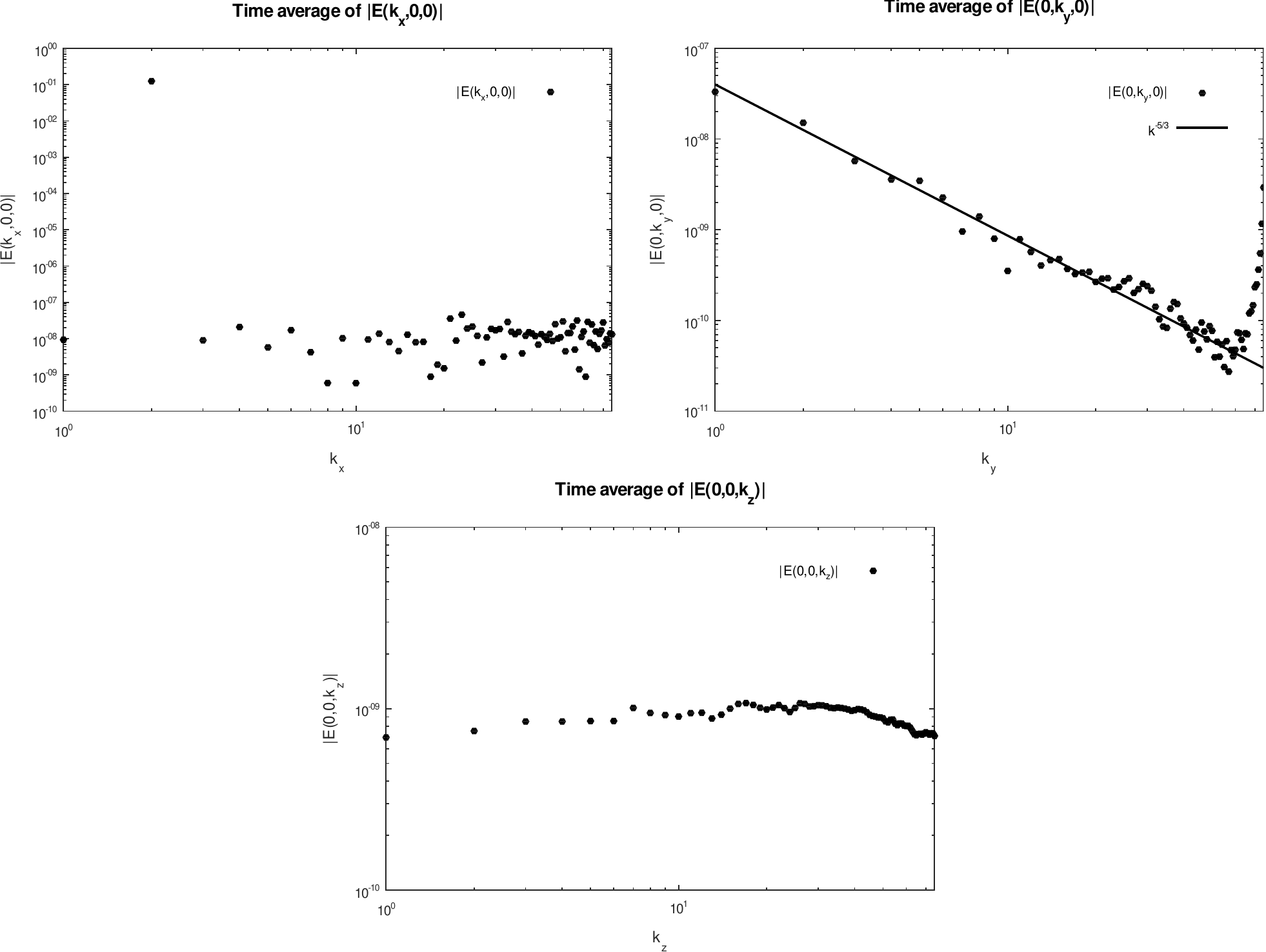}
\caption{The values of $|\hat E(k_x,0,0)|$, $|\hat E(0,k_y,0)|$ and
  $|\hat E(0,0,k_z)|$ for the gravity potential forcing, averaged
  between the times $t=8$ and $t=18$. The line $k_y^{-5/3}$ is
  given for the reference.}
\label{fig:gravity_energy}
\end{figure}

Here we repeat the simulations above with the same initial condition
\eqref{eq:u_sinx} for the gravity potential, which is obtained by
inverting the sign of \eqref{eq:num_potential} with the same
parameters. As a result, the plot of our gravity potential is a
vertical mirror image of the electrostatic potential in Figure
\ref{fig:electrostatic_potential} with an identical decay of its
respective Fourier transform, and thus we do not show it in a separate
figure.

Following the same strategy as the one for the electrostatic potential
above, here we attempt to identify the secular growth regime by
numerically integrating \eqref{eq:turbulent_velocity} with the gravity
potential until time $t=30$. Unlike the electrostatic potential, the
gravity potential does not cause the exponential blow-up within the
same time frame, however, as we show in Figure
\ref{fig:gravity_norms}, the time series of the advection norm start
developing oscillations by the time $t=18$-20 units. Thus, here we
restrict the examination to the same time interval as above for the
electrostatic potential, that is, up until $t=18$ units.

In Figure \ref{fig:gravity_velocity}, we show the snapshots of $|\hat
u_x(0,k_y,0)|$ and $|\hat u_y(k_x,0,0)|$ for the gravity potential
forcing at the same times $t=8,11,15,18$ as above for the
electrostatic potential. The general properties of the plots here are
essentially the same as those for the electrostatic potential. The top
values of the component $|\hat u_x(0,k_y,0)|$ visibly decay as
$k_y^{-5/3}$ (which corresponds to the Kolmogorov spectrum), whereas
the component $\hat u_y(k_x,0,0)$ exhibits a largely flat spectrum,
with the exception of a single large scale value which corresponds to
the shear flow \eqref{eq:u_sinx}.

In Figure \ref{fig:gravity_energy} we present the time averages of
$|\hat E(k_x,0,0)|$, $|\hat E(0,k_y,0)|$ and $|\hat E(0,0,k_z)|$ for
the gravity potential, computed between the times $t=8$ and $t=18$.
The behavior of the time-averaged energy spectra is similar to those
of the electrostatic potential above, namely, the time averages of
$|\hat E(k_x,0,0)|$ and $|\hat E(0,0,k_z)|$ have largely flat spectra,
while the time-average of $|\hat E(0,k_y,0)|$ aligns very well with
the reference line $k_y^{-5/3}$.

\subsection{Thomas--Fermi potential}

After observing the energy spectra for the electrostatic and
gravitational potentials (which are the long-range potentials), it is
interesting to look at what happens for some common short-range
interatomic potentials. Here we examine the behavior of the flow for
the same initial condition in \eqref{eq:u_sinx}, forced by the
Thomas--Fermi potential with the Bohr screening function.

\begin{figure}
\includegraphics[width=\textwidth]{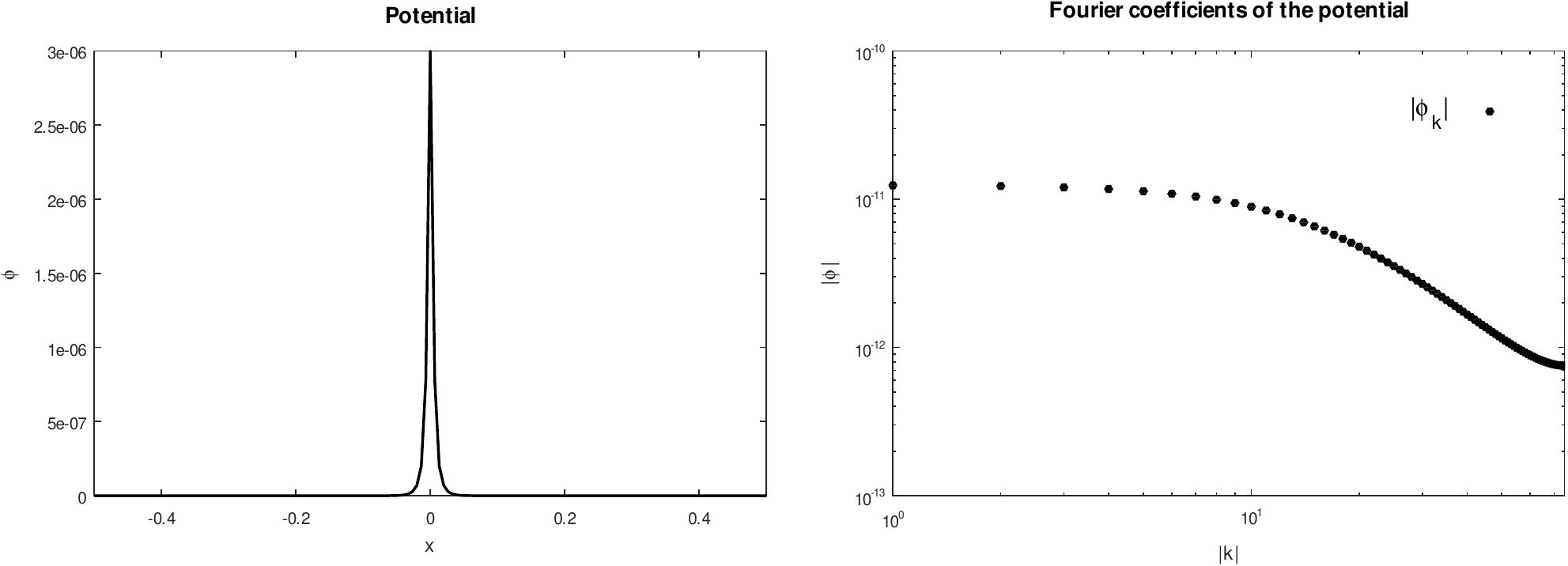}
\caption{The graph of the regularized Thomas--Fermi potential with the
  Bohr screening function.}
\label{fig:bohr_potential}
\end{figure}

\begin{figure}
\includegraphics[width=0.5\textwidth]{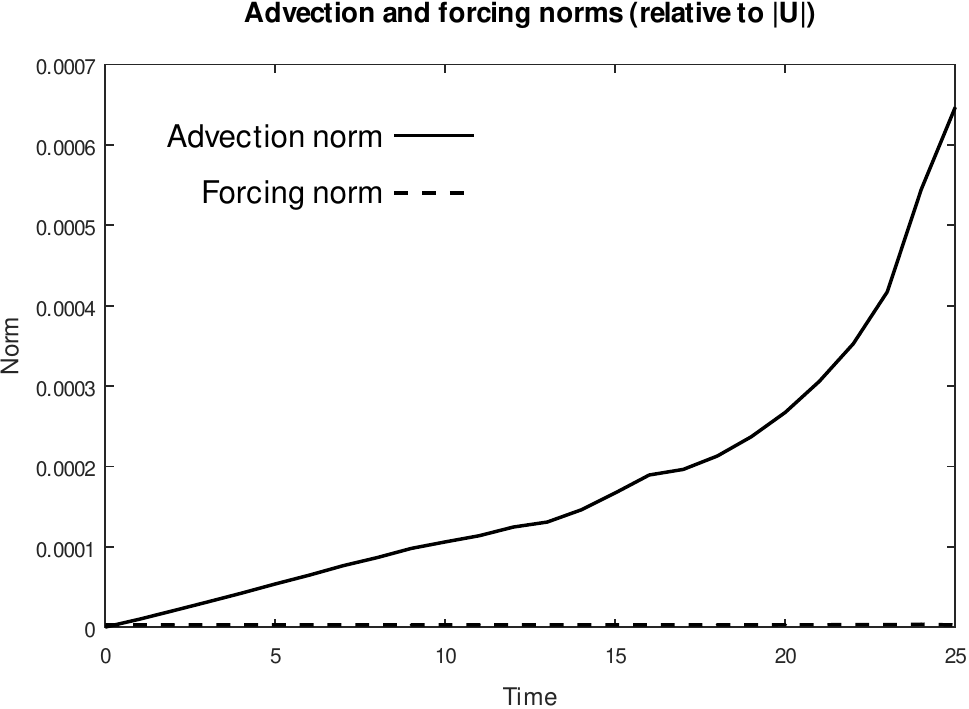}
\caption{The time series of the quadratic norms of the advection part
  of~\eqref{eq:turbulent_velocity} (solid line), and its potential
  forcing part (dashed line), for the Thomas--Fermi potential. Both
  norms are given relative to the norm of the turbulent velocity $\BV
  u$ itself.}
\label{fig:bohr_norms}
\end{figure}

\begin{figure}
\includegraphics[width=\textwidth]{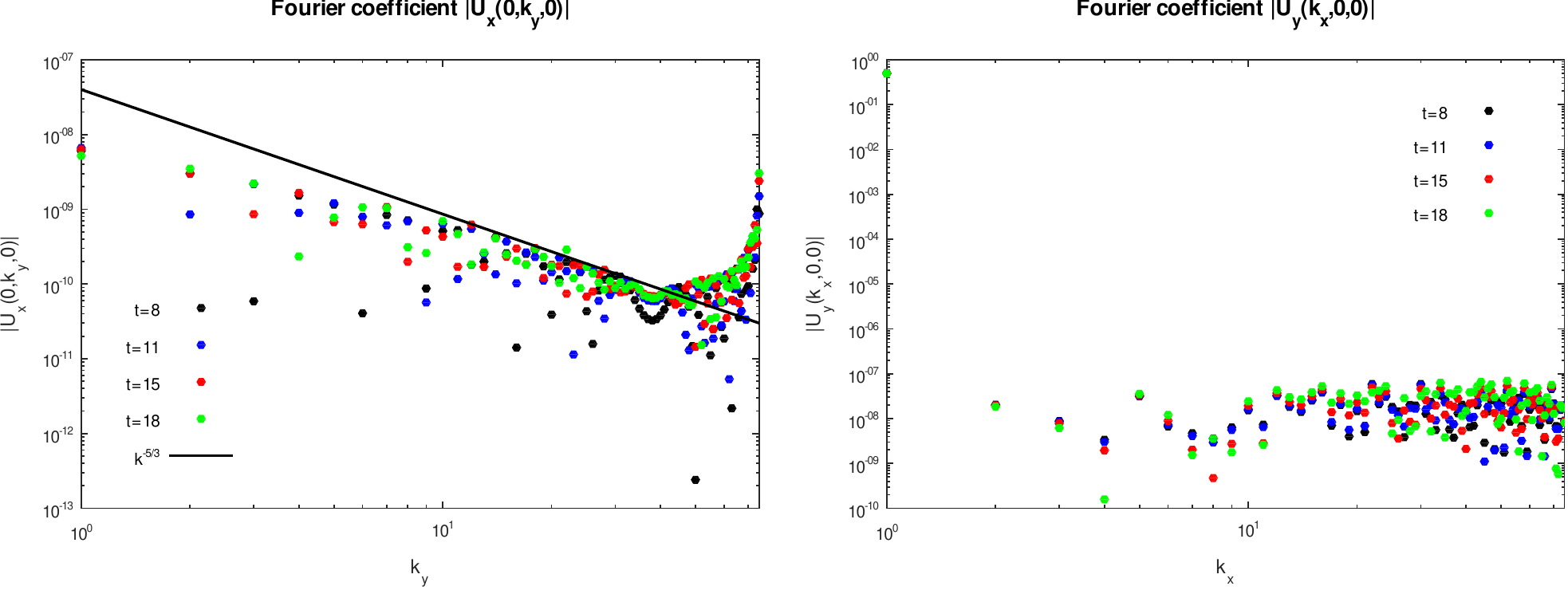}
\caption{The values of $|\hat u_x(0,k_y,0)|$ and $|\hat u_y(k_x,0,0)|$
  for the Thomas--Fermi potential forcing, captured at times
  $t=8,11,15,18$.  The line $k_y^{-5/3}$ is given for the
  reference.}
\label{fig:bohr_velocity}
\end{figure}

\begin{figure}
\includegraphics[width=\textwidth]{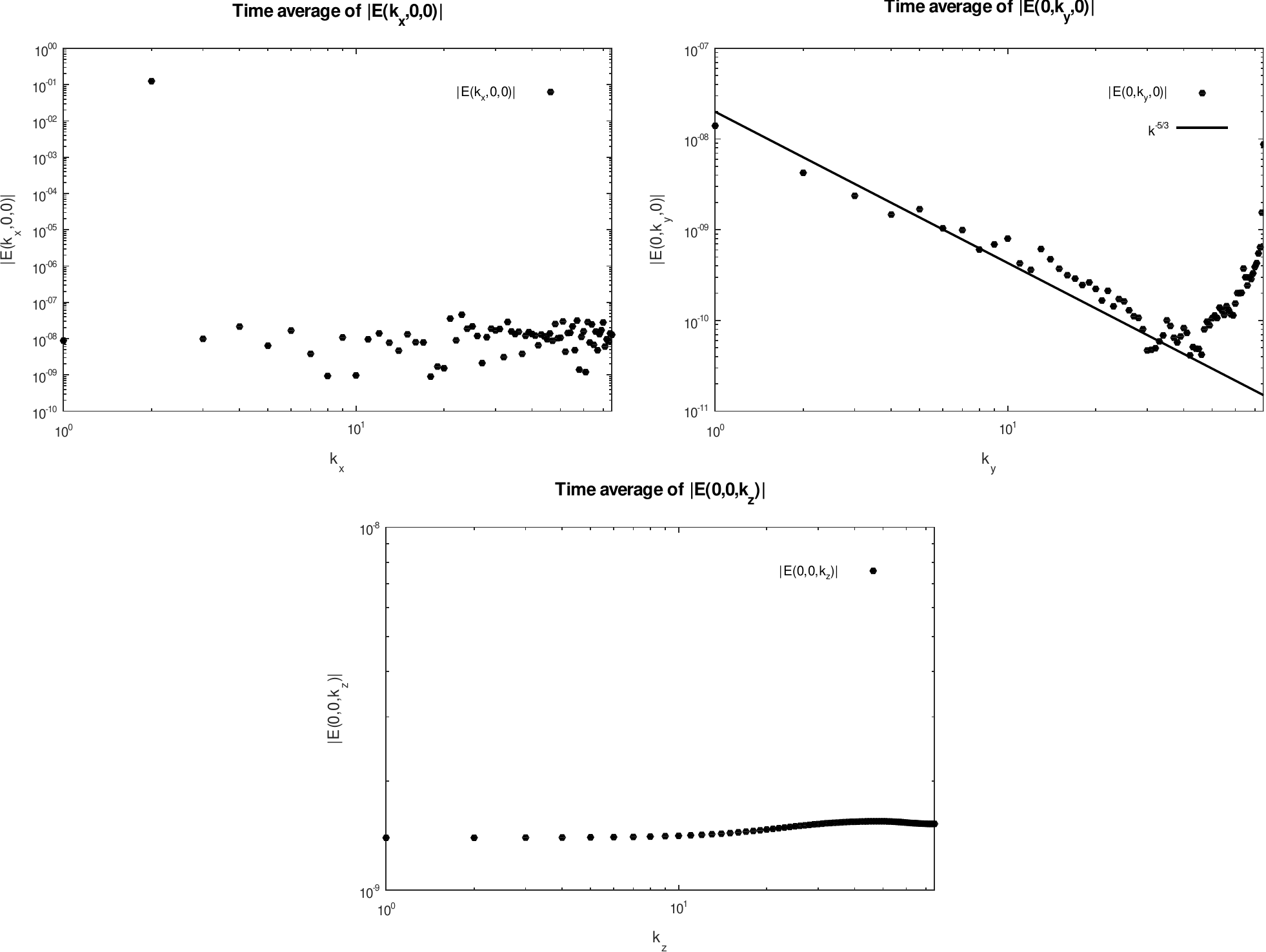}
\caption{The values of $|\hat E(k_x,0,0)|$, $|\hat E(0,k_y,0)|$ and
  $|\hat E(0,0,k_z)|$ for the Thomas--Fermi potential forcing,
  averaged between the times $t=8$ and $t=18$. The line
  $k_y^{-5/3}$ is given for the reference.}
\label{fig:bohr_energy}
\end{figure}

The Thomas--Fermi potential \citep{Tho,Fer} is given via
\begin{equation}
\label{eq:thomas_fermi}
\phi(r)=\frac{\phi_0}r\eta\left(\frac r\sigma\right),
\end{equation}
where $\sigma$ is the Bohr radius of the potential, while $\eta(r)$ is
the screening function. According to the Thomas--Fermi theory,
$\eta(r)$ satisfies the Thomas--Fermi nonlinear differential equation,
whose solution cannot be expressed in terms of elementary functions
\citep{Som}. Thus, it is not unusual to choose $\eta(r)$ empirically
instead (for example, by fitting to the scattering measurements, which
results in the Ziegler--Biersack--Littmark potential
\citep{ZieBieLit}).  Arguably, the simplest approximate screening
function for the Thomas--Fermi potential was suggested by Bohr:
\begin{equation}
\label{eq:bohr_screening}
\eta(r)=e^{-r}.
\end{equation}
Since, in the current work, it is not our goal to accurately model the
interactions between atoms, but rather to study the energy spectra for
various potential types, here we study the properties of the solution
of \eqref{eq:turbulent_velocity} with the same initial condition
\eqref{eq:u_sinx}, forced by the Thomas--Fermi potential
\eqref{eq:thomas_fermi} with the Bohr screening function
\eqref{eq:bohr_screening}. The potential is regularized in the same
way as in \eqref{eq:reg_potential}, that is, by capping it with the
inverted parabola to avoid the singularity at zero. The parameters
$\phi_0$ and $\sigma$ are chosen as follows:
\begin{equation}
\phi_0=10^{-8},\qquad\sigma=10^{-2},
\end{equation}
that is, the Bohr radius is two orders of magnitude smaller than the
size of the domain. The resulting potential, together with its Fourier
transform, is displayed in Figure \ref{fig:bohr_potential}. As we can
see, the effective range of the Thomas--Fermi potential is visibly
much shorter than that of the electrostatic and gravitational
potentials, which is also confirmed by the relatively flat spectrum of
its Fourier transform at large scales.

In this scenario, the numerical blow-up of the solution of
\eqref{eq:turbulent_velocity} occurs shortly after the time $t=25$
units.  In Figure \ref{fig:bohr_norms}, we show the time series of the
advection and forcing norms up until the blow-up time. Observe that
the linear growth of the advection norm ceases around the time $t=18$
units, and is replaced with an exponential growth leading to the
blow-up, similarly to what happened for the electrostatic potential in
Figure \ref{fig:electrostatic_norms}.

In Figure \ref{fig:bohr_velocity}, we show the snapshots of $|\hat
u_x(0,k_y,0)|$ and $|\hat u_y(k_x,0,0)|$ for the Thomas--Fermi
potential forcing at the same times $t=8,11,15,18$ as above for the
electrostatic potential. Observe that the shape of the Fourier
transform of the potential appears to affect the top values of the
component $|\hat u_x(0,k_y,0)|$ -- namely, the large scale components
of the Fourier transform fall below the $k_y^{-5/3}$ line, while the
small scale components are still aligned with it. The component $\hat
u_y(k_x,0,0)$ behaves in the same manner as before for the
electrostatic and gravity potentials, exhibiting a largely flat
spectrum, with the exception of a single large scale value which
corresponds to the shear flow \eqref{eq:u_sinx}.

In Figure \ref{fig:bohr_energy} we present the time averages of $|\hat
E(k_x,0,0)|$, $|\hat E(0,k_y,0)|$ and $|\hat E(0,0,k_z)|$ for the
Thomas--Fermi potential, computed between the times $t=8$ and $t=18$.
The behavior of the time-averaged energy spectra $|\hat E(k_x,0,0)|$
and $|\hat E(0,0,k_z)|$ is the same as that for the electrostatic and
gravity potentials, that is, they are largely flat. Surprisingly, the
time-average of $|\hat E(0,k_y,0)|$ exhibits something resembling a
``tiered'' structure, where snippets of the spectrum visibly align
with the reference line $k_y^{-5/3}$, with abrupt transitions in
between. This is, however, different from what was observed for the
electrostatic and gravity potentials, where almost the whole spectrum,
except for very small scales, aligned consistently with the reference
line.

\subsection{Lennard-Jones potential}

The second short-range interaction potential we examine here is an
analog of the Lennard-Jones interatomic potential \citep{Len}. The
chief difference between the Lennard-Jones potential and the
Thomas--Fermi potential is that the former combines the attraction at
longer distances with the repulsion at short distances.

Just as all preceding potentials, the Lennard-Jones potential has a
singularity at zero. However, unlike the preceding potentials, the
behavior of the Lennard-Jones potential leading to the singularity is
much steeper ($r^{-12}$ versus $r^{-1}$), and, unfortunately,
regularizing it with the inverted parabola, like we did above for the
electrostatic, gravity and Thomas--Fermi potentials, does not work.

\begin{figure}
\includegraphics[width=\textwidth]{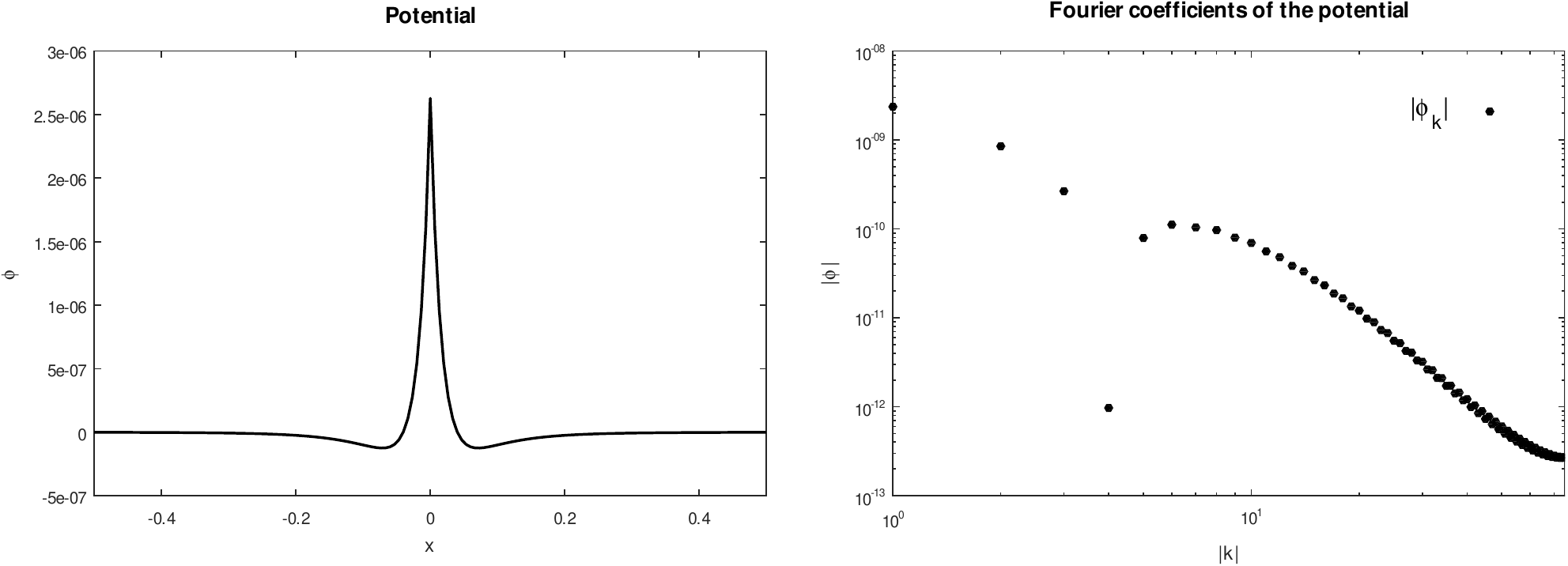}
\caption{The graph of the model Lennard-Jones potential
  $\phi^{num}_{LJ}(|x|)$ (left), and its Fourier transform
  $\hat\phi^{num}_{LJ}(\|\BV k\|)$ (right).}
\label{fig:LJ_potential}
\end{figure}

\begin{figure}
\includegraphics[width=0.5\textwidth]{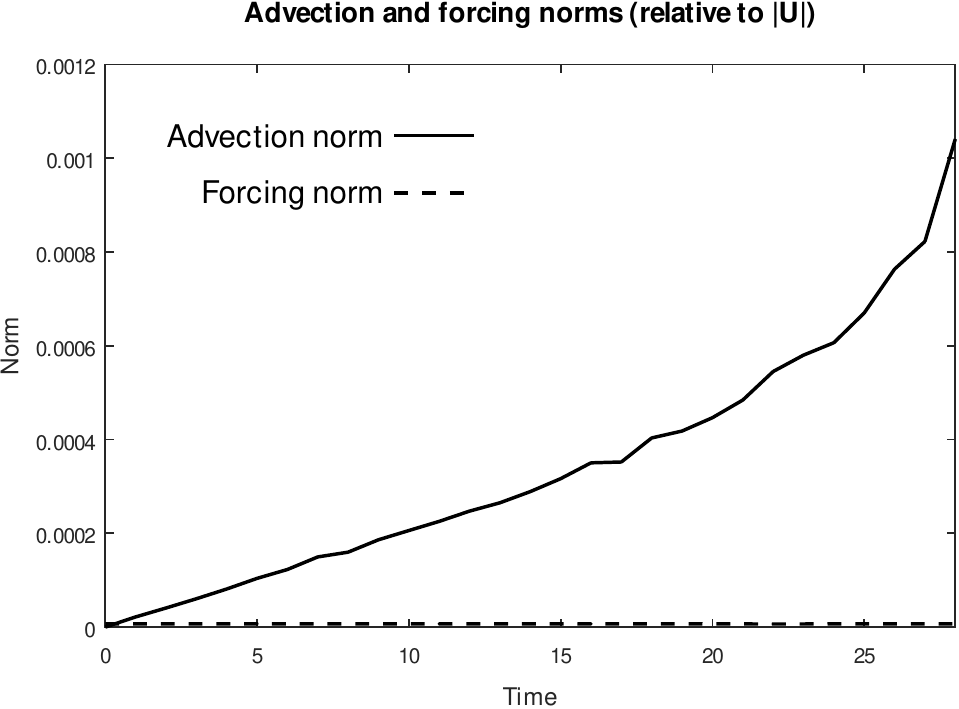}
\caption{The time series of the quadratic norms of the advection part
  of~\eqref{eq:turbulent_velocity} (solid line), and its potential
  forcing part (dashed line), for the Lennard-Jones potential. Both
  norms are given relative to the norm of the turbulent velocity $\BV
  u$ itself.}
\label{fig:LJ_norms}
\end{figure}

\begin{figure}
\includegraphics[width=\textwidth]{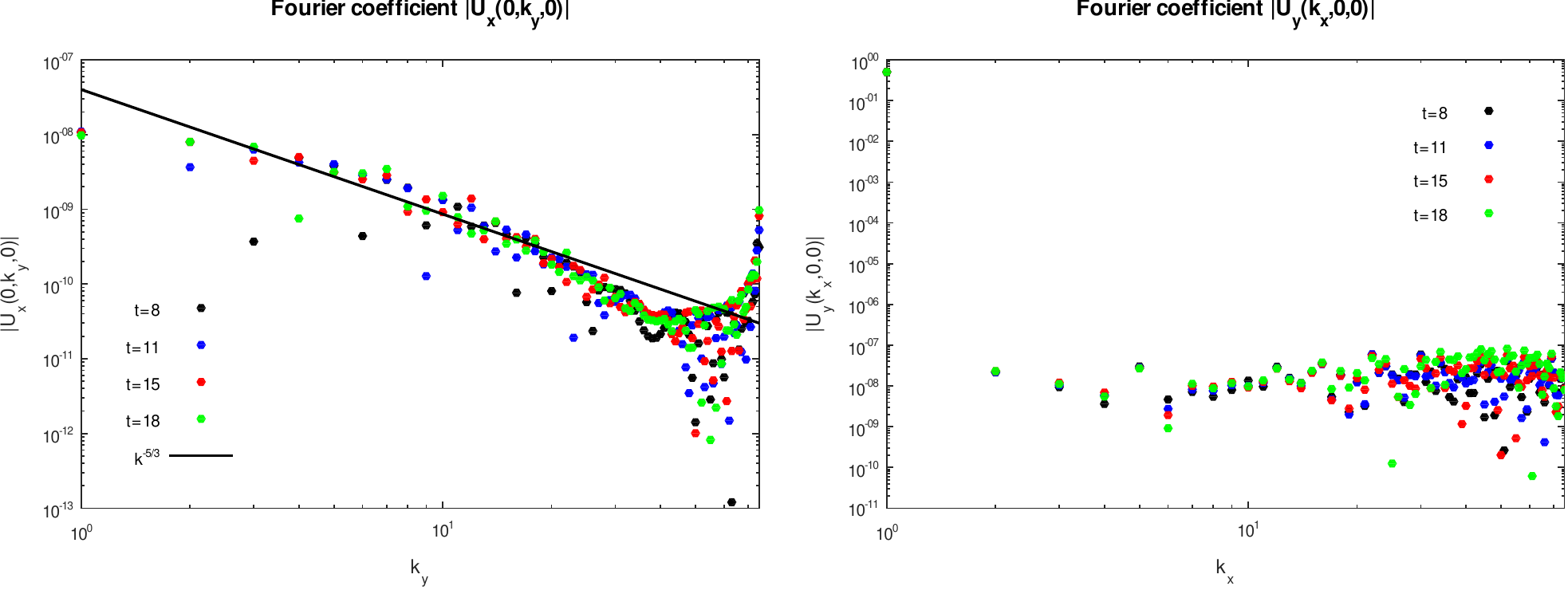}
\caption{The values of $|\hat u_x(0,k_y,0)|$ and $|\hat u_y(k_x,0,0)|$
  for the Lennard-Jones potential forcing, captured at times
  $t=8,11,15,18$.  The line $k_y^{-5/3}$ is given for the
  reference.}
\label{fig:LJ_velocity}
\end{figure}

\begin{figure}
\includegraphics[width=\textwidth]{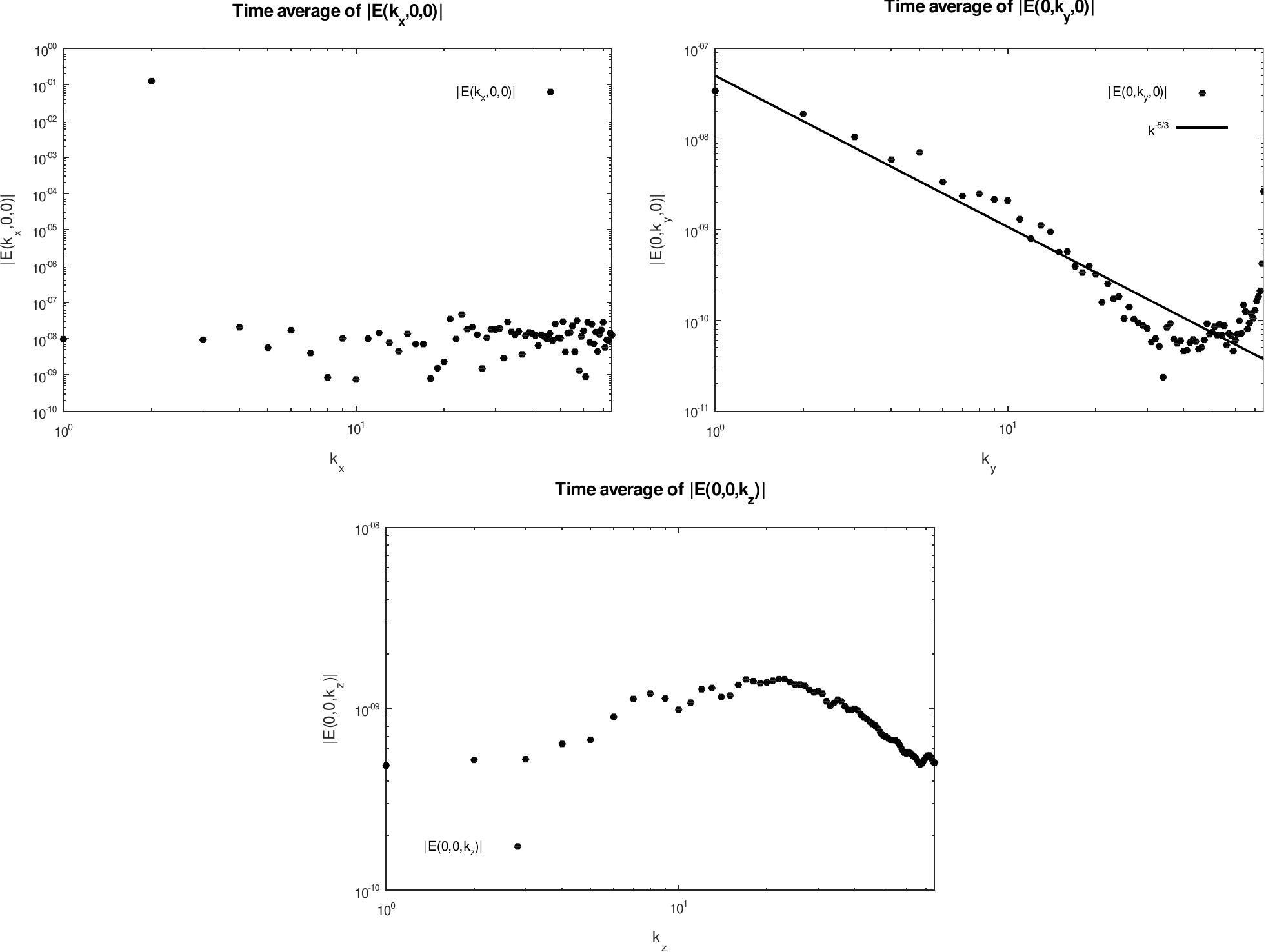}
\caption{The values of $|\hat E(k_x,0,0)|$, $|\hat E(0,k_y,0)|$ and
  $|\hat E(0,0,k_z)|$ for the Lennard-Jones potential forcing,
  averaged between the times $t=8$ and $t=18$. The line
  $k_y^{-5/3}$ is given for the reference.}
\label{fig:LJ_energy}
\end{figure}

Instead, we eliminate the singularity at zero by shifting the argument
of the potential by a positive offset, so that the infinite value is
never reached for a nonnegative argument. Namely, recall that the
Lennard-Jones potential is given via
\begin{equation}
\phi_{LJ}(r)=\phi_0\bigg[\left(\frac\sigma r\right)^{12}
  -\left(\frac\sigma r\right)^6\bigg].
\end{equation}
For the numerical simulations, we modify the above expression as
follows:
\begin{equation}
\phi_{LJ}^{num}(r)=\phi_{LJ}(r+r_0),\qquad r_0>0.
\end{equation}
The potential thus becomes shifted to the left along the horizontal
axis, such that the infinity is achieved for a negative value of the
argument. However, since the distance between the particles cannot be
negative, the expression above is finite at any model discretization
point.

For the simulation, we use the following values of the parameters:
\begin{equation}
\phi_0=10^{-6},\qquad\sigma=0.25,\qquad r_0=0.21.
\end{equation}
The graph of the model potential, together with its Fourier transform,
is shown in Figure~\ref{fig:LJ_potential}. Observe that the model
potential indeed combines the attraction at longer distances with the
repulsion at short distances, just as intended.

For the Lennard-Jones potential, the numerical blow-up of the solution
of \eqref{eq:turbulent_velocity} occurs shortly after the time $t=28$
units.  In Figure \ref{fig:LJ_norms}, we show the time series of the
advection and forcing norms of the numerical solution of
\eqref{eq:turbulent_velocity} for the Lennard-Jones potential up until
the blow-up time. Observe that the linear growth of the advection norm
ceases around the time $t=20$ units, and is replaced with an
exponential growth leading to the blow-up.

In Figure \ref{fig:LJ_velocity}, we show the snapshots of $|\hat
u_x(0,k_y,0)|$ and $|\hat u_y(k_x,0,0)|$ for the Lennard-Jones
potential forcing at the same times $t=8,11,15,18$ as for all of the
preceding cases. Observe that the shape of the Fourier transform of
the potential also affects the top values of the component $|\hat
u_x(0,k_y,0)|$ (just as for the Thomas--Fermi potential above) --
namely, the large scale components of the Fourier transform fall below
the $k_y^{-5/3}$ line, while the small scale components are still
aligned with it. The component $\hat u_y(k_x,0,0)$ behaves in the same
manner as before for the electrostatic, gravity and Thomas--Fermi
potentials, exhibiting a largely flat spectrum, with the exception of
a single large scale value which corresponds to the shear flow
\eqref{eq:u_sinx}.

In Figure \ref{fig:LJ_energy} we present the time averages of $|\hat
E(k_x,0,0)|$, $|\hat E(0,k_y,0)|$ and $|\hat E(0,0,k_z)|$ for the
Lennard-Jones potential, computed between the times $t=8$ and $t=18$.
The behavior of the time-averaged energy spectra $|\hat E(k_x,0,0)|$
and $|\hat E(0,0,k_z)|$ resembles that for the electrostatic, gravity
and Thomas--Fermi potentials, that is, the spectra do not exhibit any
systematic power decay. The scaling of time-average of $|\hat
E(0,k_y,0)|$, on the other hand, clearly follows the reference line
$k_y^{-5/3}$.

\subsection{Large scale potential (Vlasov equation)}

\begin{figure}
\includegraphics[width=0.5\textwidth]{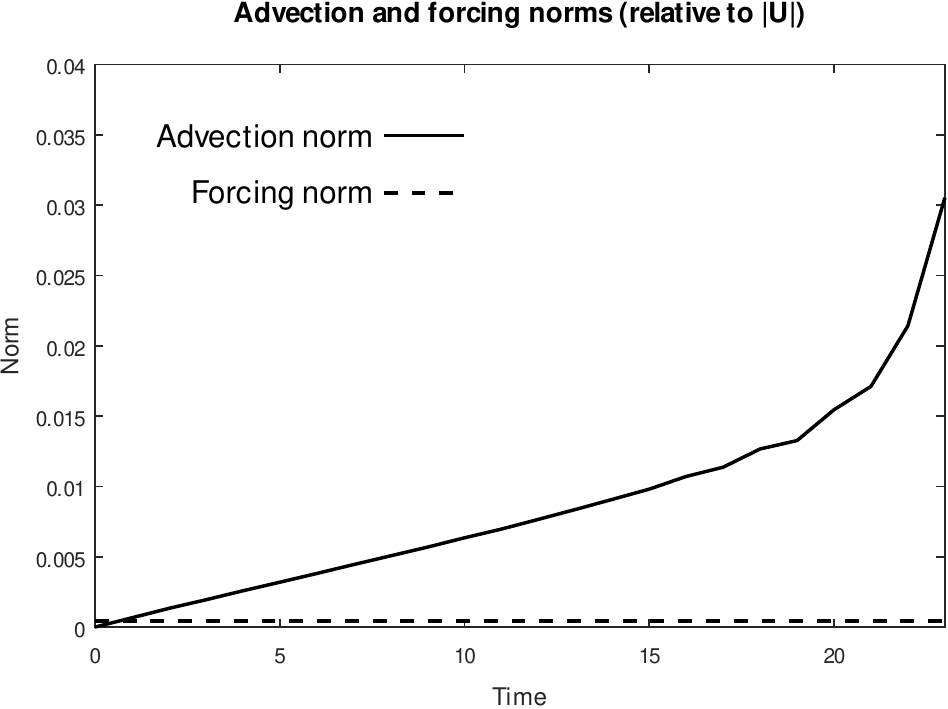}
\caption{The time series of the quadratic norms of the advection part
  of~\eqref{eq:turbulent_velocity} (solid line), and its potential
  forcing part (dashed line), for the large scale potential. Both
  norms are given relative to the norm of the turbulent velocity $\BV
  u$ itself.}
\label{fig:largescale_norms}
\end{figure}

\begin{figure}
\includegraphics[width=\textwidth]{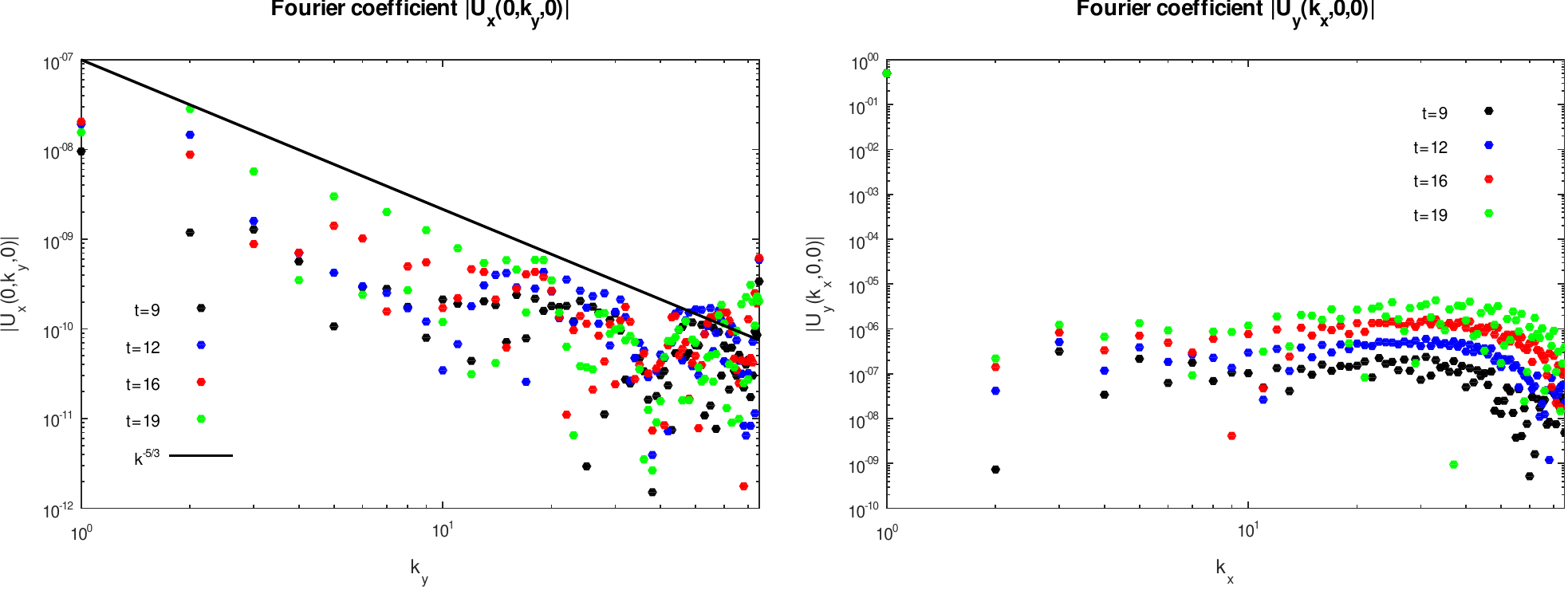}
\caption{The values of $|\hat u_x(0,k_y,0)|$ and $|\hat u_y(k_x,0,0)|$
  for the large scale potential forcing, captured at times
  $t=9,12,16,19$.  The line $k_y^{-5/3}$ is given for the
  reference.}
\label{fig:largescale_velocity}
\end{figure}

\begin{figure}
\includegraphics[width=\textwidth]{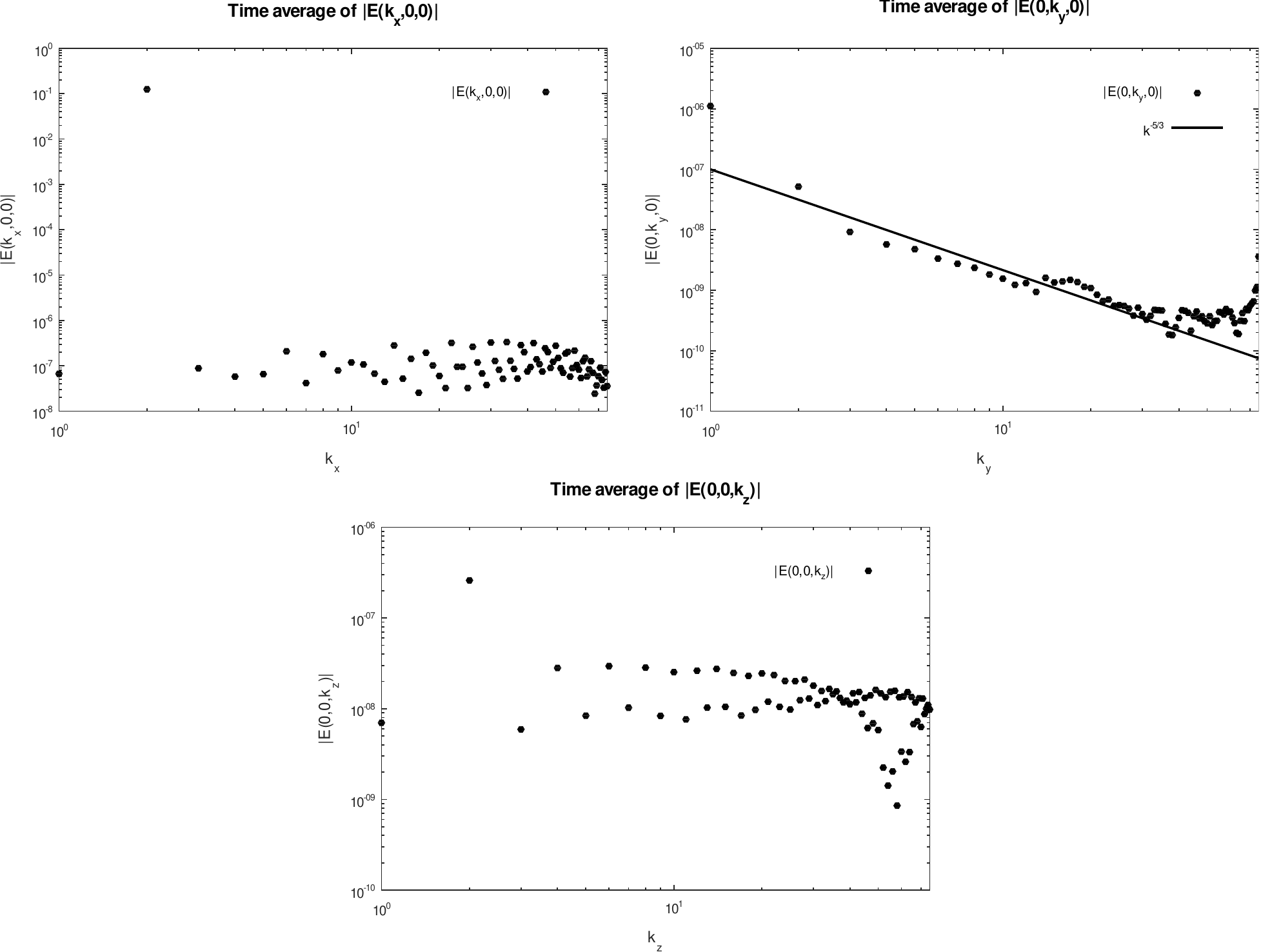}
\caption{The values of $|\hat E(k_x,0,0)|$, $|\hat E(0,k_y,0)|$ and
  $|\hat E(0,0,k_z)|$ for the large scale potential forcing, averaged
  between the times $t=9$ and $t=19$. The line $k_y^{-5/3}$ is
  given for the reference.}
\label{fig:largescale_energy}
\end{figure}

So far, we examined the dynamics of the turbulent velocity equation
\eqref{eq:turbulent_velocity} for a short-range interatomic potential
(such as Thomas--Fermi or Lennard-Jones), or a ``rangeless''
potential, such as electrostatic or gravitational. In all cases, we
observed the kinetic energy spectra consistent with the Kolmogorov
power decay. The only scenario which remains unexamined is the one
where the potential is confined solely to large scales, in the context
of the Vlasov equation \eqref{eq:vlasov} for a single particle.

In the current work, we examine our hypothesis of formation of
turbulence in \eqref{eq:turbulent_velocity} by potentials of different
types, and it is not our goal to accurately reproduce the actual
compressible gas flow with proper relations between the density,
velocity, and other variables. Therefore, here it suffices to set the
potential $\phi$ in \eqref{eq:turbulent_velocity} to a simple,
artificial combination of the large scale periodic functions:
\begin{equation}
\phi(x,y,z)=\phi_0\cos(2\pi x)\sin(2\pi y)\cos(2\pi z).
\end{equation}
For our simulation, we set $\phi_0=10^{-5}$, which yields the secular
rate of growth similar to the potentials tested above. In this
scenario, the numerical blow-up occurs shortly after the time $t=24$
units. In Figure \ref{fig:largescale_norms}, we show the time series
of the advection and forcing norms of the numerical solution of
\eqref{eq:turbulent_velocity} for the large scale potential up until
the blow-up time. Observe that the linear growth of the advection norm
ceases around the time $t=20$ units, and is replaced with the
exponential growth leading to the blow-up.

In Figure \ref{fig:largescale_velocity}, we show the snapshots of
$|\hat u_x(0,k_y,0)|$ and $|\hat u_y(k_x,0,0)|$ for the large scale
potential forcing at the times $t=9,12,16,19$. Here, the velocity
Fourier transform $|\hat u_x(0,k_y,0)|$ follows the $k_y^{-5/3}$
reference line rather poorly, as one can discern only the general
``bulk'' trend corresponding to the Kolmogorov decay, but not the
``sharp top'' as for the electrostatic or gravity potentials
above. The component $\hat u_y(k_x,0,0)$ behaves in the same manner as
before for the electrostatic, gravity, Thomas--Fermi and Lennard-Jones
potentials, exhibiting a largely flat spectrum, with the exception of
a single large scale value which corresponds to the shear flow
\eqref{eq:u_sinx}.

In Figure \ref{fig:largescale_energy} we present the time averages of
$|\hat E(k_x,0,0)|$, $|\hat E(0,k_y,0)|$ and $|\hat E(0,0,k_z)|$ for
the large scale potential, computed between the times $t=9$ and
$t=19$. The behavior of the time-averaged energy spectra $|\hat
E(k_x,0,0)|$ and $|\hat E(0,0,k_z)|$ is the same as that for the
electrostatic, gravity, Thomas--Fermi and Lennard-Jones potentials,
that is, they are largely flat. Remarkably, the scaling of the
time-average of $|\hat E(0,k_y,0)|$ follows the reference line
$k_y^{-5/3}$ almost as accurately as for the electrostatic and
gravitational potentials, and notably better than for the
Thomas--Fermi and Lennard-Jones potentials.

Concluding this section, we note that the bulk decay properties of the
kinetic energy spectrum along the direction of the large scale shear
flow depend rather weakly on the type of the potential overall, and
tend to support the Kolmogorov spectrum hypothesis for the whole
variety of studied potentials.

\section{Summary}
\label{sec:summary}

In the current work, we study the ability of the potential interaction
between particles to form turbulent structures with power decay
spectra from an initially laminar shear flow. We start with a simple
model consisting of only two particles, which interact via a
potential.  We then change the variables to those which quantify the
motion of the center of mass of the system (the mean flow), and the
difference of the coordinates of the particles (the turbulent
variables), and formulate the Liouville equation for the turbulent
coordinate and velocity variables. Alternatively, we formulate the
Vlasov equation for one of the particles in the pair, by excluding the
other particle via a simple closure.

Observing that these Liouville and Vlasov equations have the same form
(differing only in the type of the forcing potential), we derive the
hierarchy of the velocity moment transport equations for either of the
two in the same manner as in the conventional fluid mechanics. Due to
the fact that this hierarchy lacks the Boltzmann collision integral
(which is replaced by the potential forcing), we introduce a novel
closure, based upon the condition of a high Reynolds number of the
flow.  Our closure leads to a standalone equation for the velocity
variable, forced by the interaction potential.

As the turbulent velocity equation is a nonlinear second-order PDE, we
study the behavior of its solutions via numerical simulations, using a
large scale laminar shear flow as the initial condition. We examine
the resulting dynamics for the following interaction potentials:
electrostatic, gravitational, Thomas--Fermi, Lennard-Jones, as well as
the Vlasov-type large scale mean field potential. In each scenario, we
discover the regime of secular growth which precludes the exponential
blow-up, where the latter apparently occurs due to the lack of
dissipation. In all scenarios, the time-averaged kinetic energy of the
flow in this secular growth regime decays as the negative five-thirds
power of its Fourier wavenumber, which corresponds to the Kolmogorov
turbulence spectrum.

While the initial examination of the formation of turbulence,
according to our hypothesis, has been completed in the present work,
the properties of turbulence dissipation are still inaccessible in the
current state of our model. The reason for this is that, in its
present form, the turbulent velocity equation itself lacks
dissipation.  This, in turn, stems from the fact that there is no
dissipation in the simple two-particle model we study here -- the sole
interaction present in the system occurs via a potential, and is,
therefore, fully time-reversible. In order to introduce dissipation, a
promising approach seems to be to take the multiparticle system as a
starting point, and treat the collisions of the particles beyond the
first two as irreversible \citep{Abr17}. We will explore this approach
in the future work.

\ack This work was supported by the Simons Foundation grant \#636144.

\appendix

\section{Multiparticle dynamics}
\label{app:multiparticle}

Here, we consider a dynamical system which consists of $N$ identical
particles, interacting via a potential $\phi(r)$. Denoting the
coordinate and velocity of the $i$-th particle via $\BV x_i$ and $\BV
v_i$, respectively, we have the following system of equations of
motion:
\begin{equation}
\label{eq:dyn_sys_N}
\deriv{\BV x_i}t=\BV v_i,\qquad\deriv{\BV v_i}t=-\sum_{\myatop{j=1}{
    j\neq i}}^N\parderiv{}{\BV x_i}\phi(\|\BV x_i-\BV x_j\|).
\end{equation}
The total momentum and energy of all particles are preserved by the
dynamics:
\begin{equation}
\sum_{i=1}^N\BV v_i=\text{const},\qquad\sum_{i=1}^N\frac{\|\BV v\|^2}
2+\sum_{i=1}^{N-1}\sum_{j=i+1}^N\phi(\|\BV x_i-\BV x_j\|)=
\text{const}.
\end{equation}
Observe that, for a given value of the momentum, it is always possible
to choose the inertial reference frame in which the momentum becomes
zero; thus, without much loss of generality, we will further assume
that the total momentum of the system is zero.

Following our work \citep{Abr20}, we concatenate the coordinates as
$\BV X=(\BV x_1,\ldots,\BV x_N)$, and velocities as $\BV V=(\BV
v_1,\ldots,\BV v_N)$. In these notations, we can write
\begin{equation}
\deriv{\BV X}t=\BV V,\qquad\deriv{\BV V}t=-\parderiv\Phi{\BV X},\qquad
\Phi(\BV X)=\sum_{i=1}^{N-1}\sum_{j=i+1}^N\phi(\|\BV x_i-\BV x_j\|).
\end{equation}
In the variables $\BV X$ and $\BV V$, the conservation of the energy
can be expressed via
\begin{equation}
\|\BV V\|^2+2\Phi(\BV X)=\text{const}.
\end{equation}
Let $F(t,\BV X,\BV V)$ be the density of states of the dynamical
system above. Then, the Liouville equation for $F$ is given via
\begin{equation}
\label{eq:liouville_N}
\parderiv Ft+\BV V\cdot\parderiv F{\BV X}=\parderiv\Phi{\BV X}\cdot
\parderiv F{\BV V}.
\end{equation}
Any suitable $F_0$ of the form
\begin{equation}
F_0(\BV X,\BV V)=F_0\big(\|\BV V\|^2+2\Phi(\BV X)\big)
\end{equation}
is a steady state for \eqref{eq:liouville_N}. Among those, the
canonical Gibbs state is
\begin{equation}
\label{eq:F_G_N}
F_G(\BV X,\BV V)=\frac
1{(2\pi\theta_0)^{3N/2}Z_N}\exp\left(-\frac{\|\BV V\|^2+2\Phi(\BV X)}{
2\theta_0}\right),\qquad Z_N=\int e^{-\Phi(\BV X)/\theta_0}\dif\BV X,
\end{equation}
where $\theta_0$ is the kinetic temperature. The conservation of the
R\'enyi divergences is shown in the same manner as in Section
\ref{sec:particle_dynamics}; indeed, for $\psi_1(F)$ and $\psi_2(F)$
we have
\begin{multline}
\parderiv{}t\int\psi_1(F)\psi_2(F_0)\dif\BV X\dif\BV V=\int\psi_2(
F_0)\parderiv{\psi_1(F)}t\dif\BV X\dif\BV V=\int\psi_2(F_0)
\bigg(\parderiv\Phi{\BV X}\cdot\parderiv{\psi_1(F)}{\BV V}-\\-\BV
V\cdot\parderiv{\psi_1(F)}{\BV X}\bigg) \dif\BV X\dif\BV
V=\int\psi_1(F)\bigg(\BV V\cdot\parderiv{\psi_2(F_0)}{\BV
  X}-\parderiv\Phi{\BV X}\cdot \parderiv{\psi_2(F_0)}{\BV
  V}\bigg)\dif\BV X\dif\BV V=0.
\end{multline}

\subsection{Marginal distributions of the Gibbs state}

For the particles $1,\ldots,k$, $k\leq N$, we have
\begin{multline}
F^{(1,\ldots,k)}_G=\int F_G(\BV X,\BV V)\dif\BV x_{k+1}\dif\BV v_{k+1}
\ldots\dif\BV x_N\dif\BV v_N=\\=\frac 1{(2\pi\theta_0)^{3k/2}}e^{
  -(\|\BV v_1\|^2+\ldots+\|\BV v_k\|^2)/2\theta_0}\prod_{i=1}^{k-1}
\prod_{j=i+1}^k e^{-\phi(\|\BV x_i-\BV x_j\|)/\theta_0} Y^{(k)}(\BV
x_1, \ldots,\BV x_k),
\end{multline}
where $Y^{(k)}_N$ is a multiple of the $k$-particle cavity
distribution function \citep{Bou06}:
\begin{multline}
\label{eq:cavity_DF}
Y^{(k)}_N(\BV x_1,\ldots,\BV x_k)=\frac 1{Z_N}\int\bigg(\prod_{i=k+1
}^N e^{-(\phi(\|\BV x_1-\BV x_i\|)+\ldots+\phi(\|\BV x_k-\BV x_i\|))/
  \theta_0}\\\prod_{j=i+1}^N e^{-\phi(\|\BV x_i-\BV x_j\|)/\theta_0}
\bigg)\dif\BV x_{k+1}\ldots\dif\BV x_N.
\end{multline}
For the two and three particles, we can write their marginal
distributions $F_G^{(1,2)}$ and $F_G^{(1,2,3)}$ via
\begin{subequations}
\begin{multline}
F_G^{(1,2)}=\frac 1{(2\pi\theta_0)^{3/2}V}e^{-\|\BV v_1\|^2/2
  \theta_0}\frac 1{(2\pi\theta_0)^{3/2}V}e^{-\|\BV
  v_2\|^2/2\theta_0}e^{-\phi(\|\BV x_2-\BV x_1\|)/\theta_0}\\
V^2Y^{(2)}(\BV x_1,\BV x_2)=F_G^{(1)}(\BV v_1)F_G^{(2)}(\BV
v_2)e^{-\phi(\|\BV x_2-\BV x_1\|)/\theta_0}V^2 Y^{(2)}(\BV x_1,\BV
x_2),
\end{multline}
\begin{multline}
F_G^{(1,2,3)}=\frac 1{(2\pi\theta_0)^3}e^{-(\|\BV v_1\|^2+\|\BV v_2
  \|^2)/2\theta_0}\frac 1{Z_2}e^{-\phi(\|\BV x_2-\BV x_1\|)/\theta_0}
\frac 1{(2\pi\theta_0)^{3/2}V}e^{-\|\BV v_3\|^2 /2\theta_0}\\Z_2V
e^{-(\phi(\|\BV x_3-\BV x_1\|)+\phi(\|\BV x_3-\BV x_2\|) )/\theta_0}
Y^{(3)}_N(\BV x_1,\BV x_2,\BV x_3)=\\= F_G^{(1,2)}(\BV x_1,\BV v_1,\BV
x_2,\BV v_2)F_G^{(3)}(\BV v_3) e^{-(\phi(\|\BV x_3-\BV x_1\|)
  +\phi(\|\BV x_3-\BV x_2\|)) /\theta_0} Z_2V Y^{(3)}_N(\BV x_1,\BV
x_2,\BV x_3).
\end{multline}
\end{subequations}
If the gas is dilute (that is, at average distances the particles are
weakly affected by the potential interaction), then $V^2 Y^{(2)}\to
1$, $Z_2V Y^{(3)}\to 1$, and $F_G^{(1,2)}$, $F_G^{(1,2,3)}$ become
\begin{subequations}
\begin{equation}
\label{eq:F_G_12}
F_G^{(1,2)}=F_G^{(1)}(\BV v_1)F_G^{(2)}(\BV v_2) e^{-\phi(\|\BV
  x_2-\BV x_1\|)/\theta_0},
\end{equation}
\begin{equation}
\label{eq:F_G_123}
F_G^{(1,2,3)}=F_G^{(1,2)}(\BV x_1,\BV v_1,\BV x_2,\BV v_2)
F_G^{(3)}(\BV v_3) e^{-(\phi(\|\BV x_3-\BV x_1\|)+\phi(\|\BV x_3-\BV
  x_2\|)) /\theta_0}.
\end{equation}
\end{subequations}

\section{The closure for a single particle (Vlasov equation)}
\label{app:vlasov}

Here, we isolate a single particle (say, \#1), and examine the
transport of its marginal distribution $F^{(1)}$, given via
\begin{equation}
F^{(1)}(t,\BV x_1,\BV v_1)=\int F(t,\BV X,\BV V)\dif \BV x_2\dif\BV
v_2\ldots\dif\BV x_N\dif\BV v_N.
\end{equation}
Integrating the Liouville equation in \eqref{eq:liouville_N} over all
particles but the first one, in the absence of boundary effects we
arrive at
\begin{equation}
\parderiv{F^{(1)}}t+\BV v_1\cdot\parderiv{F^{(1)}}{\BV x_1}
=\sum_{i=2}^N\int\parderiv{}{\BV x_1}\phi(\|\BV y-\BV x_1\|) \cdot
\parderiv{}{\BV v_1}F^{(1,i)}(\BV x_1,\BV v_1,\BV y,\BV w)\dif\BV
y\dif\BV w.
\end{equation}
Assuming that the gas is dilute, and the state $F$ is close to the
Gibbs equilibrium, we use the same closure for $F^{(1,i)}$ as in
\eqref{eq:F_G_12}:
\begin{equation}
F^{(1,i)}(\BV x_1,\BV v_1,\BV y,\BV w)=F^{(1)}(\BV x_1,\BV
v_1)F^{(i)}(\BV y,\BV w) e^{-\phi(\|\BV y-\BV x_1\|)/\theta_0}.
\end{equation}
This leads to
\begin{multline}
\int\parderiv{}{\BV x_1}\phi(\|\BV y-\BV x_1\|) \cdot \parderiv{}{\BV
  v_1}F^{(1,i)}(\BV x_1,\BV v_1,\BV y,\BV w)\dif\BV y\dif\BV w=
\parderiv{}{\BV v_1}F^{(1)}(\BV x_1,\BV v_1)\cdot\\\cdot\int\parderiv{
}{\BV x_1}\phi(\|\BV y-\BV x_1\|)e^{-\phi(\|\BV y-\BV x_1\|)/\theta_0}
F^{(i)}(\BV y,\BV w)\dif\BV y\dif\BV w=\parderiv{\bar\phi_i(\BV
  x_1)}{\BV x_1}\cdot\parderiv{F^{(1)}}{\BV v_1},
\end{multline}
where
\begin{equation}
\bar\phi_i(\BV x)=-\theta_0\int e^{-\phi(\|\BV x-\BV y\|)/\theta_0}
\rho_i(\BV y)\dif\BV y,\qquad\rho_i(\BV y)=\int F^{(i)}(\BV y,\BV
w)\dif\BV w.
\end{equation}
Now, denoting
\begin{equation}
\bar\phi(\BV x)=\sum_{i=2}^N\bar\phi_i(\BV x),
\end{equation}
we arrive at the Vlasov equation
\begin{equation}
\parderiv{F^{(1)}}t+\BV v_1\cdot\parderiv{F^{(1)}}{\BV x_1}
=\parderiv{\bar\phi(\BV x_1)}{\BV x_1}\cdot\parderiv{F^{(1)}}{\BV
  v_1}.
\end{equation}
This equation has the same structure as \eqref{eq:liouville} and
\eqref{eq:vlasov}. Just as in \eqref{eq:vlasov}, the potential
$\bar\phi$ is, generally, time-dependent.

\section{The closure for a pair of particles}
\label{app:closure_pair}

Here, we isolate a pair of particles (say, \#1 and \#2), and examine
the transport of their marginal distribution $F^{(1,2)}$, given via
\begin{equation}
F^{(1,2)}(t,\BV x_1,\BV v_1,\BV x_2,\BV v_2)=\int F(t,\BV X,\BV V)\dif
\BV x_3\dif\BV v_3\ldots\dif\BV x_N\dif\BV v_N.
\end{equation}
Integrating the Liouville equation in \eqref{eq:liouville_N} over all
particles but the first two, and assuming the absence of any boundary
effects, we arrive at
\begin{equation}
\bigg(\parderiv{}t+\BV v_1\cdot\parderiv{}{\BV x_1}+\BV v_2\cdot
\parderiv{}{\BV x_2}\bigg)F^{(1,2)}=\int\bigg(\parderiv\Phi{\BV x_1}
\cdot\parderiv{}{\BV v_1}+\parderiv\Phi{\BV x_2}\cdot\parderiv{}{\BV
  v_2}\bigg)F\dif \BV x_3\dif\BV v_3\ldots\dif\BV x_N\dif\BV v_N,
\end{equation}
where the terms with the derivatives in $\BV v_i$, $i>2$, are
integrated out. Above, observe that the derivatives of the potential
can be written via
\begin{subequations}
\begin{equation}
\parderiv\Phi{\BV x_1}=-\phi'(\|\BV x_2-\BV x_1 \|)\frac{\BV x_2-\BV
  x_1}{\|\BV x_2-\BV x_1\|}+\sum_{i=3}^N \parderiv{}{\BV x_1}
\phi(\|\BV x_i-\BV x_1\|),
\end{equation}
\begin{equation}
\parderiv\Phi{\BV x_2}=\phi'(\|\BV x_2-\BV x_1 \|)\frac{\BV x_2-\BV
  x_1}{\|\BV x_2-\BV x_1\|}+\sum_{i=3}^N \parderiv{}{\BV x_2}
\phi(\|\BV x_i-\BV x_2\|),
\end{equation}
\end{subequations}
which allows to write the transport equation for $F^{(1,2)}$ via
\begin{multline}
\label{eq:BBGKY}
\bigg(\parderiv{}t+\BV v_1\cdot\parderiv{}{\BV x_1}+\BV v_2\cdot
\parderiv{}{\BV x_2}\bigg)F^{(1,2)}=\phi'(\|\BV x_2-\BV x_1\|)\frac{
  \BV x_2-\BV x_1}{\|\BV x_2-\BV x_1\|}\cdot \bigg(\parderiv{}{\BV
  v_2}-\parderiv{}{\BV v_1}\bigg)F^{(1,2)}+\\+\sum_{i=3}^N\int\bigg(
\parderiv{}{\BV x_1}\phi(\|\BV x_i-\BV x_1\|)\cdot\parderiv{}{\BV v_1}
+\parderiv{}{\BV x_2}\phi(\|\BV x_i-\BV x_2\|)\cdot\parderiv{}{\BV
  v_2}\bigg) F^{(1,2,i)}\dif\BV x_i\dif\BV v_i,
\end{multline}
where $F^{(1,2,i)}$ is the marginal distribution of the three
particles -- the first, second, and $i$-th. The equation for the
marginals in \eqref{eq:BBGKY} is a part of the
Bogoliubov--Born--Green--Kirkwood--Yvon hierarchy
\citep{Bog,BorGre,Kir}.

In order to proceed further, we need to introduce a closure for
$F^{(1,2,i)}$. In similar scenarios in the literature
\citep{Cer,CerIllPul} it is assumed that the marginal distributions
for different particles are independent; however, in the present
context such an assumption would obviously be incorrect, which is
indicated by the structure of $F_0$ in \eqref{eq:F_G_N}. Instead, here
we follow our earlier work \citep{Abr17} and assume that the structure
of $F^{(1,2,i)}$ mimics that of the corresponding three-particle
marginal of $F_G$ in \eqref{eq:F_G_123}; namely, $F^{(1,2,i)}$ has the
form
\begin{equation}
\label{eq:F_closure}
F^{(1,2,i)}(\BV x_1,\BV v_1,\BV x_2,\BV v_2,\BV x_i,\BV v_i)=
F^{(1,2)}(\BV x_1,\BV v_1,\BV x_2,\BV v_2) F^{(i)}(\BV x_i,\BV v_i)
e^{-(\phi(\|\BV x_i-\BV x_1\|)+\phi(\|\BV x_i-\BV x_2\|))/\theta_0},
\end{equation}
where $F^{(i)}$ is the single-particle marginal distribution for the
$i$-th particle. Observe that the closure in \eqref{eq:F_closure}
becomes exact if $F$ is the steady state \eqref{eq:F_G_N}. Under this
assumption, we arrive at
\begin{multline}
\sum_{i=3}^N\int\bigg(\parderiv{}{\BV x_1}\phi(\|\BV x_i-\BV x_1\|)
\cdot\parderiv{}{\BV v_1}+\parderiv{}{\BV x_2}\phi(\|\BV x_i-\BV x_2
\|)\cdot\parderiv{}{\BV v_2}\bigg)F^{(1,2,i)}\dif\BV x_i\dif\BV v_i=
\\=\bigg( \parderiv{F^{(1,2)}}{\BV v_1}\cdot\parderiv{}{\BV x_1}
+\parderiv{F^{(1,2)}}{\BV v_2}\cdot\parderiv{}{\BV x_2}\bigg)\bigg(
-(N-2) \theta_0\int e^{-(\phi(\|\BV x_1-\BV z\|)+\phi(\|\BV x_2-\BV
  z\|))/\theta_0}\bar\rho(\BV z)\dif\BV z\bigg),
\end{multline}
where we denote
\begin{equation}
\bar\rho(\BV z)=\frac 1{N-2}\sum_{i=3}^N\rho_i(\BV z).
\end{equation}
The transport equation for $F^{(1,2)}$ is now closed. Next, let us
switch to the variables $\BV x$, $\BV v$, $\BV y$ and $\BV w$ from
\eqref{eq:mean_turbulent_variables}. In these variables, the closure
term becomes
\begin{multline}
-(N-2)\theta_0\int e^{-(\phi(\|\BV x_1-\BV z\|)+\phi(\|\BV x_2-\BV
  z\|))/\theta_0} \bar\rho(\BV z)\dif\BV z=\frac{2-N}2\theta_0\int
\Big[e^{-\phi(\|\BV x_2-\BV x_1 -\BV z\|)/\theta_0}\\\bar\rho(\BV
  x_1+\BV z)+ e^{-\phi(\|\BV x_2 -\BV x_1+\BV z\|)/ \theta_0}
  \bar\rho(\BV x_2+\BV z)\Big] e^{-\phi( \|\BV z\|)/\theta_0}\dif\BV
z=\frac{2-N}2\theta_0\int \Big[e^{-\phi(\|\BV x-\BV z\|)/\theta_0}
  \\\bar\rho(\BV y-\BV x/2+\BV z)+ e^{-\phi(\|\BV x+\BV
    z\|)/\theta_0}\bar\rho(\BV y+\BV x/2+\BV z)\Big]e^{-\phi(\|\BV
  z\|)/\theta_0}\dif\BV z=\bar\phi(\BV x,\BV y),
\end{multline}
where we observe that the dependence on $\BV y$ is only present in the
arguments of $\bar\rho$ (and therefore vanishes at equilibrium, where
$\bar\rho$ becomes uniform). Recalling
\eqref{eq:mean_turbulent_derivatives}, we write the transport equation
for $F^{(1,2)}$ as
\begin{equation}
\label{eq:F_12}
\parderiv{F^{(1,2)}}t+\BV v\cdot\parderiv{F^{(1,2)}}{\BV x}+\BV w\cdot
\parderiv{F^{(1,2)}}{\BV y}=2\parderiv{(\phi+\bar\phi)}{\BV x}\cdot
\parderiv{F^{(1,2)}}{\BV v}+\frac 12\parderiv{\bar \phi}{\BV y}\cdot
\parderiv{F^{(1,2)}}{\BV w}.
\end{equation}
Here, the integration over $\dif\BV y\dif\BV w$ does not directly lead
to the closed evolution of the marginal distribution in $(\BV x,\BV
v)$, because, unlike $\phi$, $\bar\phi$ is a function of $\BV y$.
However, if the dependence of $\bar\phi$ on $\BV y$ is weak enough so
that
\begin{equation}
\int\bar\phi(\BV x,\BV y) F^{(1,2)}\dif\BV y\approx\bar\phi(\BV x)\int
F^{(1,2)}\dif\BV y,
\end{equation}
where $\bar\phi(\BV x)$ is the average of $\bar\phi(\BV x,\BV y)$ over
the second argument, then the integration of \eqref{eq:F_12} in
$\dif\BV y\dif\BV w$ leads to \eqref{eq:liouville}, with the forcing
potential given via $\phi(\|\BV x\|)+\bar\phi(\BV x)$.


\begin{thebibliography}{33}
\providecommand{\natexlab}[1]{#1}
\providecommand{\url}[1]{\texttt{#1}}
\expandafter\ifx\csname urlstyle\endcsname\relax
  \providecommand{\doi}[1]{doi: #1}\else
  \providecommand{\doi}{doi: \begingroup \urlstyle{rm}\Url}\fi

\bibitem[Abramov(2017)]{Abr13}
R.V. Abramov.
\newblock Diffusive {B}oltzmann equation, its fluid dynamics, {C}ouette flow
  and {K}nudsen layers.
\newblock \emph{Physica A}, 484:\penalty0 532--557, 2017.

\bibitem[Abramov(2019)]{Abr17}
R.V. Abramov.
\newblock The random gas of hard spheres.
\newblock \emph{J}, 2\penalty0 (2):\penalty0 162--205, 2019.

\bibitem[Abramov(2020)]{Abr20}
R.V. Abramov.
\newblock Turbulent energy spectrum via an interaction potential.
\newblock \emph{J. Nonlinear Sci.}, 30:\penalty0 3057--3087, 2020.

\bibitem[Abramov and Otto(2018)]{AbrOtt}
R.V. Abramov and J.T. Otto.
\newblock Nonequilibrium diffusive gas dynamics: {P}oiseuille microflow.
\newblock \emph{Physica D}, 371:\penalty0 13--27, 2018.

\bibitem[Batchelor(2000)]{Bat}
G.K. Batchelor.
\newblock \emph{An Introduction to Fluid Dynamics}.
\newblock Cambridge University Press, New York, 2000.

\bibitem[Bogoliubov(1946)]{Bog}
N.N. Bogoliubov.
\newblock Kinetic equations.
\newblock \emph{J. Exp. Theor. Phys.}, 16\penalty0 (8):\penalty0 691--702,
  1946.

\bibitem[Boltzmann(1872)]{Bol}
L.~Boltzmann.
\newblock Weitere {S}tudien \"uber das {W}\"armegleichgewicht unter
  {G}asmolek\"ulen.
\newblock \emph{Sitz.-Ber. Kais. Akad. Wiss. (II)}, 66:\penalty0 275--370,
  1872.

\bibitem[Born and Green(1946)]{BorGre}
M.~Born and H.S. Green.
\newblock A general kinetic theory of liquids {I}: The molecular distribution
  functions.
\newblock \emph{Proc. Roy. Soc. A}, 188:\penalty0 10--18, 1946.

\bibitem[Boubl\'ik(1986)]{Bou06}
T.~Boubl\'ik.
\newblock Background correlation functions in the hard sphere systems.
\newblock \emph{Mol. Phys.}, 59\penalty0 (4):\penalty0 775--793, 1986.

\bibitem[Buchhave and Velte(2017)]{BucVel}
P.~Buchhave and C.M. Velte.
\newblock Measurement of turbulent spatial structure and kinetic energy
  spectrum by exact temporal-to-spatial mapping.
\newblock \emph{Phys. Fluids}, 29\penalty0 (8):\penalty0 085109, 2017.

\bibitem[Cercignani(1975)]{Cer}
C.~Cercignani.
\newblock \emph{Theory and Application of the {B}oltzmann Equation}.
\newblock Elsevier Science, New York, 1975.

\bibitem[Cercignani et~al.(1994)Cercignani, Illner, and Pulvirenti]{CerIllPul}
C.~Cercignani, R.~Illner, and M.~Pulvirenti.
\newblock The mathematical theory of dilute gases.
\newblock In \emph{Applied Mathematical Sciences}, volume 106. Springer-Verlag,
  1994.

\bibitem[Fermi(1927)]{Fer}
E.~Fermi.
\newblock A statistical method for determining some properties of the atom.
\newblock \emph{Rend. Accad. Naz. Lincei}, 6:\penalty0 602--607, 1927.

\bibitem[Golse(2005)]{Gols}
F.~Golse.
\newblock \emph{The {B}oltzmann Equation and its Hydrodynamic Limits}, volume~2
  of \emph{Handbook of Differential Equations: Evolutionary Equations},
  chapter~3, pages 159--301.
\newblock Elsevier, 2005.

\bibitem[Grad(1949)]{Gra}
H.~Grad.
\newblock On the kinetic theory of rarefied gases.
\newblock \emph{Comm. Pure Appl. Math.}, 2\penalty0 (4):\penalty0 331--407,
  1949.

\bibitem[Kirkwood(1946)]{Kir}
J.G. Kirkwood.
\newblock The statistical mechanical theory of transport processes {I}: General
  theory.
\newblock \emph{J. Chem. Phys.}, 14:\penalty0 180--201, 1946.

\bibitem[Kolmogorov(1941{\natexlab{a}})]{Kol41a}
A.N. Kolmogorov.
\newblock Local structure of turbulence in an incompressible fluid at very high
  {R}eynolds numbers.
\newblock \emph{Dokl. Akad. Nauk SSSR}, 30:\penalty0 299--303,
  1941{\natexlab{a}}.

\bibitem[Kolmogorov(1941{\natexlab{b}})]{Kol41b}
A.N. Kolmogorov.
\newblock Decay of isotropic turbulence in an incompressible viscous fluid.
\newblock \emph{Dokl. Akad. Nauk SSSR}, 31:\penalty0 538--541,
  1941{\natexlab{b}}.

\bibitem[Kolmogorov(1941{\natexlab{c}})]{Kol41c}
A.N. Kolmogorov.
\newblock Energy dissipation in locally isotropic turbulence.
\newblock \emph{Dokl. Akad. Nauk SSSR}, 32:\penalty0 19--21,
  1941{\natexlab{c}}.

\bibitem[Kullback and Leibler(1951)]{KulLei}
S.~Kullback and R.~Leibler.
\newblock On information and sufficiency.
\newblock \emph{Ann. Math. Stat.}, 22:\penalty0 79--86, 1951.

\bibitem[Lennard-Jones(1924)]{Len}
J.E. Lennard-Jones.
\newblock On the determination of molecular fields. -- {II}. {F}rom the
  equation of state of a gas.
\newblock \emph{Proc. R. Soc. Lond. A}, 106\penalty0 (738):\penalty0 463--477,
  1924.

\bibitem[Nastrom and Gage(1985)]{NasGag}
G.D. Nastrom and K.S. Gage.
\newblock A climatology of atmospheric wavenumber spectra of wind and
  temperature observed by commercial aircraft.
\newblock \emph{J. Atmos. Sci.}, 42\penalty0 (9):\penalty0 950--960, 1985.

\bibitem[Obukhov(1941)]{Obu41}
A.M. Obukhov.
\newblock On the distribution of energy in the spectrum of a turbulent flow.
\newblock \emph{Izv. Akad. Nauk SSSR Ser. Geogr. Geofiz.}, 5:\penalty0
  453--466, 1941.

\bibitem[Obukhov(1949)]{Obu49}
A.M. Obukhov.
\newblock Structure of the temperature field in turbulent flow.
\newblock \emph{Izv. Akad. Nauk SSSR Ser. Geogr. Geofiz.}, 13:\penalty0 58--69,
  1949.

\bibitem[Obukhov(1962)]{Obu62}
A.M. Obukhov.
\newblock Some specific features of atmospheric turbulence.
\newblock \emph{J. Geophys. Res.}, 67\penalty0 (8):\penalty0 3011--3014, 1962.

\bibitem[R\'enyi(1961)]{Ren}
A.~R\'enyi.
\newblock On measures of entropy and information.
\newblock In \emph{Proceedings of the Fourth Berkeley Symposium on Mathematical
  Statistics and Probability, Volume 1: Contributions to the Theory of
  Statistics}, pages 547--561, Berkeley, CA, 1961. University of California
  Press.

\bibitem[Reynolds(1895)]{Rey}
O.~Reynolds.
\newblock On the dynamical theory of incompressible viscous fluids and the
  determination of the criterion.
\newblock \emph{Phil. Trans. Roy. Soc. A}, 186:\penalty0 123--164, 1895.

\bibitem[Sommerfeld(1932)]{Som}
A.~Sommerfeld.
\newblock Asymptotic integration of the {T}homas--{F}ermi differential
  equation.
\newblock \emph{Rend. Accad. Naz. Lincei}, 15:\penalty0 788--792, 1932.

\bibitem[Struchtrup and Torrilhon(2003)]{StruTor}
H.~Struchtrup and M.~Torrilhon.
\newblock Regularization of {G}rad's 13-moment equations: Derivation and linear
  analysis.
\newblock \emph{Phys. Fluids}, 15:\penalty0 2668--2680, 2003.

\bibitem[Thomas(1927)]{Tho}
L.H. Thomas.
\newblock The calculation of atomic fields.
\newblock \emph{Proc. Camb. Phil. Soc.}, 23\penalty0 (5):\penalty0 542--548,
  1927.

\bibitem[Vlasov(1938)]{Vla}
A.A. Vlasov.
\newblock On vibration properties of electron gas.
\newblock \emph{J. Exp. Theor. Phys.}, 8\penalty0 (3):\penalty0 291, 1938.

\bibitem[Weller et~al.(1998)Weller, Tabor, Jasak, and Fureby]{WelTabJasFur}
H.G. Weller, G.~Tabor, H.~Jasak, and C.~Fureby.
\newblock A tensorial approach to computational continuum mechanics using
  object-oriented techniques.
\newblock \emph{Computers in Physics}, 12\penalty0 (6):\penalty0 620--631,
  1998.

\bibitem[Ziegler et~al.(1985)Ziegler, Biersack, and Littmark]{ZieBieLit}
J.F. Ziegler, J.P. Biersack, and U.~Littmark.
\newblock \emph{The Stopping and Range of Ions in Solids}, volume~1 of
  \emph{Stopping and Ranges of Ions in Matter}.
\newblock Pergamon, 1985.

\end{thebibliography}
\end{document}